%% file: main.tex
\documentclass[sigconf]{acmart}

\input{preambles/macros}

\AtBeginDocument{%
  }

\setcopyright{none}
\renewcommand\footnotetextcopyrightpermission[1]{%
  \footnotetext{$^{*}$Both authors contributed equally to this paper.}}

\begin{document}

\title{\X: An Extensible Cross-Platform Point-Based Differentiable Renderer}

\author{
{Steve Rhyner$^{{\scriptscriptstyle *}1,2}$}\quad\hspace{0.25em}
{Sankeerth Durvasula$^{{\scriptscriptstyle *}1,2}$}\quad\hspace{0.25em}
{Aleksandr Kovalev$^{1}$}\quad\hspace{0.25em}
{Hansel Jia$^{1}$}\quad\hspace{0.25em}
{Adrian Zhao$^{1,2}$}\\
{Mrutunjayya Mrutunjayya$^{3}$}\quad\hspace{0.5em}
{Nilesh Ahuja$^{3}$}\quad\hspace{0.5em}
{Selvakumar Panneer$^{3}$}\\
{Christina Giannoula$^{4}$}\quad\hspace{0.5em}
{Nandita Vijaykumar$^{1,2}$}\vspace{0.7em}\\
{\small$^1$University of Toronto}
\quad
{\small$^2$Vector Institute}
\quad
{\small$^3$Intel}
\quad
{\small$^4$Max Planck Institute for Software Systems}
}

\renewcommand{\shortauthors}{Rhyner, et al.}

\begin{abstract}
Point-based differentiable rendering underpins modern 3D reconstruction, novel-view synthesis, and learning-based graphics pipelines, but developing new rendering methods often requires extensive low-level implementation, hardware-specific kernels, and manually written backward passes. This limits rapid prototyping, reproducibility, exploration, and deployment, especially across diverse hardware platforms. This paper presents XPR, an extensible cross-platform framework for point-based differentiable rendering. XPR introduces a high-level programming interface that separates method-specific logic from the shared rendering pipeline, allowing users to implement new methods in a few lines of code. Its pipeline decomposes rendering into modular, statically shaped parallel operations that can be lowered by a cross-platform compiler to GPUs, TPUs, CPUs, and other ML accelerators. We demonstrate implementations of 3DGS, 3DGUT, and LinPrim, with only a few 100s lines of Python code, each of which can be compiled to a range of hardware platforms with the XLA compiler. These results show that XPR enables fast experimentation and portable execution for emerging point-based differentiable rendering systems.

\end{abstract}

\maketitle
\pagestyle{fancy}
\fancyhead{}
\fancyfoot[C]{\thepage}
\renewcommand{\headrulewidth}{0pt}

\input{sections/Intro_and_Motivation}
\input{sections/Background}
\input{sections/Approach}
\input{sections/Detailed_Design}

\input{sections/Methodology}

\input{sections/Evaluation}

\input{sections/Discussion}
\input{sections/Conclusion}

\bibliographystyle{ACM-Reference-Format}
\bibliography{references/refs}

\appendix
\input{sections/Appendix}

\end{document}

%% file: preambles/macros.tex
\usepackage{booktabs}
\usepackage{multirow}
\usepackage{xspace}
\usepackage{listings}
\usepackage{xcolor}
\usepackage{subcaption}
\usepackage{cleveref}
\usepackage{tikz}
\usepackage{enumitem}
\usepackage{colortbl}

\usepackage{amsmath}
\usepackage{textcomp}
\usepackage{placeins}
\usepackage{float}

\usepackage{array}
\usepackage{pifont, makecell, graphicx, wasysym}

\usepackage{algorithm}
\usepackage{algpseudocode}
\usepackage{bm}
\usepackage{url}
\usepackage{tabularx}
\newcolumntype{Y}{>{\raggedright\arraybackslash}X}
\newcolumntype{M}{>{\centering\arraybackslash}p{1.5em}}

\definecolor{lstBackground}{HTML}{F7F8FA}
\definecolor{lstBorder}{HTML}{C8D1DA}
\definecolor{lstKeyword}{HTML}{005CC5}
\definecolor{lstBuiltin}{HTML}{6F42C1}
\definecolor{lstConstant}{HTML}{D73A49}
\definecolor{lstComment}{HTML}{22863A}
\definecolor{lstString}{HTML}{E36209}
\definecolor{lstNumber}{HTML}{6A737D}
\definecolor{lstXPR}{HTML}{B3186D}

\lstdefinestyle{python}{
    language=Python,
    basicstyle=\ttfamily\scriptsize,
    backgroundcolor=\color{lstBackground},
    keywordstyle=\color{lstKeyword}\bfseries,
    keywordstyle=[2]\color{lstBuiltin},
    keywordstyle=[3]\color{lstXPR}\bfseries,
    commentstyle=\color{lstComment}\itshape,
    stringstyle=\color{lstString},
    deletekeywords=[2]{float, abs, max,min},
    morekeywords={def,return},
    morekeywords=[2]{world_2_camera_projection_matrix,get_R,compute_sigma_points,world_2_cam,compute_mu_2D,approximate_2D_gaussian,lowpass_filter,compute_conic_2D,mip_splatting_2D,compute_extents,compute_radius,get_bounding_box,cam_2_world,compute_max_response_canonical,compute_slab_data,compute_color,slab_m_at_corners,max_over_4_corners,min_over_pairs,slab_intersect,compute_cov_2D,compute_point_with_max_response},
    morekeywords=[3]{project,tile_cull,pixel_info,evaluate,MethodSpec,PrimitiveParams,ProjectResult,EvaluateResult,PixelInfoResult,XPR_3DGS,XPR_3DGUT,XPR_LinPrim},
    numbers=left,
    numberstyle=\tiny\color{lstNumber},
    numbersep=11pt,
    xleftmargin=15pt,
    xrightmargin=5pt,
    linewidth=\linewidth,
    breaklines=true,
    breakatwhitespace=true,
    postbreak=\mbox{\textcolor{lstNumber}{$\hookrightarrow$}\space},
    showstringspaces=false,
    columns=fullflexible,
    keepspaces=true,
    tabsize=2,
    upquote=true,
    aboveskip=4pt,
    belowskip=2pt,
    lineskip=-1pt,
    frame=single,
    framerule=0.4pt,
    framesep=6pt,
    rulecolor=\color{lstBorder},
    captionpos=b,
    escapeinside={(*@}{@*)},
}

\newcommand{\pmark}{\LEFTcircle}
\newcommand{\rowsep}{\specialrule{0.2pt}{0pt}{0pt}}
\newcommand{\X}[0]{\textcolor{black}{XPR}\xspace}

\newcommand{\secref}[1]{§\ref{#1}}

\newcommand{\operation}[0]{\textcolor{black}{module}\xspace}
\newcommand{\operations}[0]{\textcolor{black}{modules}\xspace}

\newcommand{\gatherop}[0]{\texttt{\textcolor[HTML]{2C7BB6}{gather}}\xspace}
\newcommand{\scatterop}[0]{\texttt{\textcolor[HTML]{27AE60}{scatter}}\xspace}
\newcommand{\prefixsumop}[0]{\texttt{\textcolor[HTML]{E66101}{prefix-sum}}\xspace}

\newcommand*\circled[1]{\tikz[baseline=(char.base)]{
    \node[shape=circle, fill, inner sep=0pt, minimum size=2.2ex]
      (char) {\textcolor{white}{\scalebox{0.85}{\sffamily\bfseries #1}}};}}

\newcommand{\cmark}{\textcolor{green!60!black}{\ding{51}}} 
\newcommand{\xmark}{\textcolor{red}{\ding{55}}}

\usepackage{ragged2e}
\usepackage[font=small,labelfont=bf,labelsep=period]{caption}

%% file: sections/Intro_and_Motivation.tex
\section{Introduction and Motivation}
\label{sec:introduction}

Differentiable rendering has become a core tool for recovering 3D scenes from images, with applications in 3D reconstruction~\cite{mildenhall2021nerf}, novel view synthesis~\cite{kerbl20233d}, medical imaging~\cite{cai2024radiative}, and 3D asset generation~\cite{poole2022dreamfusion}. Among differentiable rendering methods, point-based scene representation methods, such as 3D Gaussian Splatting (3DGS)~\cite{kerbl20233d}, model 3D scenes using spatial primitives defined by a 3D point, color, and density, and have become dominant due to their ability to render high-quality images at interactive frame rates. 

Differentiable rendering techniques and their applications have become an active area of research and development in academia and industry. Since 3DGS~\cite{kerbl20233d}, point-based rendering has evolved into a rapidly expanding design space, with recent works exploring new primitive types~\cite{von2025linprim, held2026triangle}, projection and pixel color/feature computation~\cite{wu20253dgut}, rendered features such as depth~\cite{zhang2026rade} and semantics~\cite{qin2024langsplat}, rasterization strategies~\cite{radl2024stopthepop}, and optimizations that improve efficiency and scalability~\cite{durvasula2025contrags,qian2025c3dgs}. Point-based renderers are also increasingly used as differentiable operators inside larger learning systems, including training pipelines for generative modeling~\cite{tang2024dreamgaussian}, camera pose optimization~\cite{matsuki2024gaussian}, and robotics~\cite{lu2024manigaussian}. More recently, world-model and video-generation pipelines use differentiable rendering to explicitly enforce geometric consistency during training~\cite{parkerholder2024genie2,team2025hunyuanworld,huang2025voyager,gao2024cat3d}.

These trends expose a gap in existing point-based differentiable rendering frameworks. First, to enable \emph{quick exploration, prototyping, and code reproduction}, new methods should be easy to express by changing only the method-specific logic: the primitive representation, projection rule, pixel color evaluation, rendered features, or optimization strategy. The rest of the rendering pipeline—e.g., visibility filtering, tiling, sorting, blending, scheduling, and differentiation—should be reusable across methods. In practice, however, existing renderers often entangle these concerns in specialized low-level kernels, so each new variant requires substantial reimplementation and, in many cases, manually derived backward passes. Second, when used within a larger learning system, they inherit a \emph{portability} requirement: because gradients must flow through the renderer at every training step, the renderer must execute on the same accelerator as the rest of the pipeline. As training infrastructure expands to non-NVIDIA GPUs, Google TPUs, AWS Trainium, and other accelerators, these renderers must execute on the same accelerator and within the same compiler stack.

\input{tables/prior_work_comparison}

These limitations define two technical challenges. \textbf{(1) \emph{Extensibility.}} For ease of programming, a programmer should be able to introduce a new primitive, change how primitives are projected onto the screen, modify how a primitive contributes to a pixel's color, or add new rendered features (e.g., depth, semantics) by specifying only the new logic. None of these changes should require reworking tiling, parallelization across primitives and pixels, data layout, or the backward pass.
Existing frameworks have limitations in programming ease that are summarized in Table~\ref{tab:framework-comparison}: Most are written in CUDA or Vulkan, require thousands of lines per method, and demand manually derived backward passes. The programmer must often reason about parallelization, intermediate data movement, and tile-based scheduling (Dec). gsplat~\cite{ye2025gsplat} provides fast implementations of 3DGS on both NVIDIA and AMD GPUs, but it requires significant CUDA/HIP code changes when extending to any new method. slang-gaussian~\cite{slanggaussian} and Taichi-splatting~\cite{taichisplatting} raise the level of abstraction, but both still require significant codebase changes for a new method (Var).

\textbf{(2) \emph{Mapping parallelism across compute hardware.}}
To map a single rendering framework with high-level programming abstractions across hardware platforms, we must expose parallelism in a hardware-agnostic form that a cross-platform compiler can map to diverse architectures—threads and warps on GPUs, static-shaped tensor operations on TPUs~\cite{jouppi2017datacenter} and Trainium~\cite{aws2026neuronxtrace, aws2026neuroncc}. Most existing point-based renderers are written for specific hardware, e.g., using CUDA for NVIDIA GPUs, or HIP for AMD GPUs (summarized in Table~\ref{tab:framework-comparison}). Vulkan- and WebGPU-based systems~\cite{vkgsrepo, vk_gaussian_splatting,takimoto2022dressi, 3dgscpp, brush} broaden compatibility to consumer GPUs with graphics pipelines beyond NVIDIA, but are currently incompatible with data center GPUs as they do not support Vulkan graphics drivers. Thus, they cannot be integrated with larger ML systems, e.g., generative/world models, on non-NVIDIA data center GPUs or accelerators like the TPU and Trainium. Furthermore, Vulkan-based systems comprise low-level kernels that do not provide abstractions for point-based differentiable rendering. Taichi~\cite{hu2019taichi} supports multiple backends, but the point-based rendering framework, Taichi-splatting~\cite{taichisplatting}, requires CUDA-specific modules.

We propose \X, a point-based differentiable rendering framework designed to address both challenges. \X has three design goals: \textbf{(1) Easy to program for quick prototyping:} \X should offer an intuitive high-level programming interface. A programmer implements a new point-based rendering method by specifying only the method-specific logic in a few lines of code without modifying the rendering pipeline, its parallelization, or its data layout. \textbf{(2) General:} \X should be sufficiently expressive to cover a large  design space of point-based differentiable rendering. Specifically, we design \X for any differentiable rendering pipeline that can be expressed as a mapping between point-based primitives and pixels in an image plane. 
\textbf{(3) Portable and Interoperable:}  \X should express the pipeline's computation and  parallelism in a  high-level hardware-agnostic form that \emph{(i)} a cross-platform compiler can lower to diverse accelerators and \emph{(ii)} is compatible with diverse IRs like StableHLO~\cite{stablehlo}, and ATen IR (torch.fx)~\cite{reed2022torch} which integrate with ML frameworks.

\X is designed based on three key ideas:
\textbf{(1) A four-function interface that separates method-specific logic.} \X abstracts the rendering pipeline as a function that computes a single primitive's contribution at a single pixel. This function is exposed to the programmer as four interface functions—\texttt{project}, \texttt{tile\_cull}, \texttt{pixel\_info}, and \texttt{evaluate}—that capture only the method-specific logic, while decoupling primitive filtering, tile assignment, depth sorting, alpha-blending, parallelization, and differentiation.
\textbf{(2) Single-element functions.} Each interface function operates on a single primitive, a single pixel, or a single primitive--pixel pair. The programmer never writes loops over primitives, pixels, or their intersections—XPR automatically generates the iteration over all primitives and pixels and filters out non-contributing primitive-pixel pairs through tiling and culling. \textbf{(3) Portable decomposition of the rendering pipeline.} \X decomposes the rendering pipeline into a sequence of \emph{modules}, each with a single, explicit dimension of parallelism—e.g., across primitives, across tiles, or across pixels. Each module's computation is expressed using building blocks available across all accelerators: data parallel element-wise operations over a batch of primitives, pixels, tiles, or primitive-tile pairs, and standard parallel primitives such as prefix-sum, sort, scatter, and gather. This exposes the renderer's parallelism in a form that a cross-platform compiler can lower to threads and warps on GPUs, or to batched operations on TPUs, from a single implementation. Critically, all tensor dimensions and loop trip counts are fixed at compile time—converting the pipeline's inherently dynamic structure, such as variable per-tile primitive lists and data-dependent blend loops, into statically shaped operations that satisfy the constraints of many ML accelerators and IRs.

\input{tables/flexibility_generality}

To demonstrate the simplicity and flexibility of \X, we implement and evaluate point-based rendering methods, including 3DGS~\cite{kerbl20233d}, LinPrim~\cite{von2025linprim}, and 3DGUT~\cite{wu20253dgut} (see~\secref{sec:ease_of_implementing_rendering_methods}). We show that the implementation requires up to an order-of-magnitude fewer lines of code than its CUDA counterpart. Beyond these, Table~\ref{tab:x-generality} (non-comprehensively) lists more recent works that can be easily implemented with \X. We validate rendering quality by evaluating PSNR, SSIM, and LPIPS against ground-truth images, and find that \X matches the corresponding CUDA baselines across these metrics. When implemented with the JAX/XLA compiler, we demonstrate that \X can execute on NVIDIA, AMD, and Intel GPUs, and Google TPUs. On the NVIDIA L40S GPU, \X reaches rendering speeds that are, on average, 97.18\% of those of hand-optimized CUDA baselines.

This paper makes the following contributions. (1) We present \X, the first framework that enables expressing point-based differentiable rendering methods using intuitive high-level programming interfaces that decouple method-specific logic from rasterization, parallelization, and hardware mapping. \X generates high-level parallelized hardware-agnostic modules that can be mapped to diverse compute hardware. (2) We provide a Python-based implementation of \X interfaced with XLA that can be used for exploration, rapid prototyping, teaching, and code reproduction on a range of hardware platforms. (3) We demonstrate ease-of-programming and generality by implementing 3DGS, 3DGUT, and LinPrim in 260 to 304 lines each, matching baseline rendering quality on 17 scenes from three standard benchmarks, and report rendering speeds across four hardware targets for three methods. (4) To facilitate future research, we aim to open-source the complete codebase of \X.

%% file: tables/prior_work_comparison.tex
\begin{table}[t]
\caption{Comparison of point-based differentiable rendering systems.}
\label{tab:framework-comparison}
\centering
\scriptsize
\renewcommand{\arraystretch}{1.08}
\setlength{\tabcolsep}{1.4pt}
\resizebox{\columnwidth}{!}{%
\begin{tabular}{@{}l@{\hskip 6pt}l@{\hskip 5pt}*{11}{c}@{}}
\toprule
\textbf{System} & \textbf{Implementation} & \multicolumn{5}{c}{\textbf{Programmability}} & \multicolumn{6}{c}{\textbf{Portability}} \\
\cmidrule(lr){3-7} \cmidrule(lr){8-13}
 & & Pt & Var. & HL & Dec. & AD & NV & AMD & Int & TPU & Trn & CPU \\
\midrule
3DGS rast.~\cite{kerbl20233d} & CUDA & \cmark & \xmark & \xmark & \xmark & \xmark & \cmark & \pmark & \xmark & \xmark & \xmark & \xmark \\ \rowsep
gsplat~\cite{ye2025gsplat} & CUDA/HIP & \cmark & \pmark & \pmark & \pmark & \xmark & \cmark & \cmark & \xmark & \xmark & \xmark & \xmark \\ \rowsep
OpenSplat~\cite{opensplat} & CUDA/HIP & \cmark & \xmark & \xmark & \xmark & \xmark & \cmark & \cmark & \xmark & \xmark & \xmark & \cmark \\ \rowsep
Nvdiffrast~\cite{laine2020modular} & CUDA & \xmark & \xmark & \pmark & \pmark & \xmark & \cmark & \xmark & \xmark & \xmark & \xmark & \xmark \\ \rowsep
PyTorch3D~\cite{pytorch3d} & CUDA & \cmark & \pmark & \pmark & \pmark & \xmark & \cmark & \xmark & \xmark & \xmark & \xmark & \pmark \\ \rowsep
Kaolin~\cite{KaolinLibrary} & CUDA & \xmark & \xmark & \pmark & \xmark & \xmark & \cmark & \xmark & \xmark & \xmark & \xmark & \xmark \\ \rowsep
Mitsuba3~\cite{Mitsuba3} & Dr.Jit & \xmark & \xmark & \pmark & \pmark & \cmark & \cmark & \xmark & \xmark & \xmark & \xmark & \cmark \\ \rowsep
slang-gaussian-rast.~\cite{slanggaussian} & Slang/SLANG.D & \cmark & \xmark & \xmark & \xmark & \cmark & \cmark & \xmark & \xmark & \xmark & \xmark & \xmark \\ \rowsep
taichi-splatting~\cite{taichisplatting} & Taichi & \cmark & \xmark & \cmark & \xmark & \cmark & \cmark & \pmark & \pmark & \xmark & \xmark & \cmark \\ \rowsep
Brush~\cite{brush} & Rust/WGSL/Burn & \cmark & \xmark & \xmark & \xmark & \xmark & \cmark & \pmark & \pmark & \xmark & \xmark & \xmark \\ \rowsep
Dressi~\cite{takimoto2022dressi} & Vulkan-Compute & \xmark & \xmark & \xmark & \xmark & \cmark & \cmark & \pmark & \pmark & \xmark & \xmark & \xmark \\ \rowsep
3DGS.cpp~\cite{3dgscpp} & Vulkan-Compute & \cmark & \xmark & \xmark & \xmark & \xmark & \cmark & \pmark & \pmark & \xmark & \xmark & \xmark \\ \rowsep
vkgs~\cite{vkgsrepo} & Vulkan-Graphics & \cmark & \xmark & \xmark & \xmark & \xmark & \pmark & \pmark & \pmark & \xmark & \xmark & \xmark \\ \rowsep
vulkan\_gaussian\_splat.~\cite{vk_gaussian_splatting} & Vulkan-Graphics & \cmark & \pmark & \xmark & \xmark & \xmark & \pmark & \pmark & \pmark & \xmark & \xmark & \xmark \\
\midrule
\textbf{Ours} & JAX/XLA & \cmark & \cmark & \cmark & \cmark & \cmark & \cmark & \cmark & \cmark & \cmark & \cmark & \cmark \\
\bottomrule
\end{tabular}%
}

\vspace{2pt}
\begin{minipage}{\columnwidth}
\scriptsize
\textbf{Legend.}
Pt: \emph{point-based} rendering framework.
Var.: ability to change the primitive, projection, or evaluation rule without rewriting the renderer.
HL: high-level user code.
Dec.: method-specific renderer logic decoupled from scheduling, tiling, and hardware mapping.
AD: automatic differentiation for renderer code; hand-written backward kernels wrapped by host autograd are marked unsupported.
NV, AMD, Int, TPU, Trn, and CPU denote NVIDIA GPU, AMD GPU, Intel GPU, Google TPU, Amazon Trainium, and CPU.
\cmark: supported.
\pmark: partial or restricted.
\xmark: unsupported.
\end{minipage}
\end{table}

%% file: tables/flexibility_generality.tex
\begin{table}[t]
\caption{Generality of \X across point-based rendering methods. Each row shows a variation class expressible in \X, and representative works.}
\label{tab:x-generality}
\centering
\scriptsize
\renewcommand{\arraystretch}{1.10}
\setlength{\tabcolsep}{2.0pt}
\resizebox{\columnwidth}{!}{%
\begin{tabular}{@{}>{\raggedright\arraybackslash}p{1.95cm}>{\raggedright\arraybackslash}p{6.90cm}@{}}
\toprule
\textbf{Variation class} & \textbf{Published point-based rendering methods expressible in \X} \\
\midrule
Primitive type &
3DGS~\cite{kerbl20233d}, 2DGS~\cite{huang20242d}, GES~\cite{hamdi2024ges}, TriangleSplatting~\cite{held2026triangle}, LinPrim~\cite{von2025linprim}, EVER~\cite{mai2025ever}, 3D Superquadric Splatting~\cite{macswayne20263d} \\
\midrule
Pixel color/feature computation & 3DGS~\cite{kerbl20233d}, 2DGS~\cite{huang20242d}, GES~\cite{hamdi2024ges}, 3DGUT~\cite{wu20253dgut}, LinPrim~\cite{von2025linprim}, EVER~\cite{mai2025ever} \\
\midrule
Rendered feature & 3DGS~\cite{kerbl20233d}, LangSplat~\cite{qin2024langsplat}, LangSplatv2~\cite{li2025langsplatv2}, SplaTAM~\cite{keetha2024splatam}, GS-SLAM~\cite{yan2024gs}, Surfels~\cite{dai2024high} \\
\midrule
Primitive-tile selection & StopThePop~\cite{radl2024stopthepop}, FlashGS~\cite{feng2025flashgs} \\
\midrule
Stored representation & ContraGS~\cite{durvasula2025contrags}, C3DGS~\cite{qian2025c3dgs} \\
\bottomrule
\end{tabular}%
}
\end{table}

%% file: sections/Background.tex
\section{Background}
\label{sec:background}

We review the point-based rendering pipeline using 3DGS~\cite{kerbl20233d} as an example. Each 3DGS primitive is a 3D Gaussian density with center $\mu$ and covariance $\Sigma=RSS^\top R^\top$ (where $R$ is rotation and $S$ diagonal scale), given by $G(\mathrm{\mathbf{x}}) = \exp\bigl(-\tfrac{1}{2}(\mathrm{\mathbf{x}} - \mu)^\top \Sigma^{-1}(\mathrm{\mathbf{x}} - \mu)\bigr)$, opacity $o$, and spherical harmonics for view-dependent color. The rendering pipeline has four stages.

\noindent \textbf{Preprocess.} The preprocess stage computes, for each 3D Gaussian, a 2D footprint (2D mean and 2D covariance) on the image plane~\cite{zwicker2002ewa}, an axis-aligned bounding box in tile coordinates, and per-primitive attributes such as the tile count, depth, and view-dependent color. We use \textit{projection} to denote the transformation that, given the camera, maps a primitive's 3D position and shape parameters to its screen-space footprint. Primitives outside the view, with negligible opacity, or smaller than one tile, are flagged for filtering.

\noindent \textbf{Filter.} The filter stage removes primitives flagged during the preprocess step. Only visible primitives are kept for rasterization.

\noindent \textbf{Rasterizer.} Each visible 3D Gaussian primitive is assigned to the tiles its bounding box intersects, and the renderer builds a depth-sorted list of contributing primitives per tile. Optimizations such as tile culling~\cite{radl2024stopthepop} further prune non-contributing primitives from each list.

\noindent \textbf{Shader.} For each pixel within a tile, the renderer evaluates each 3D Gaussian primitive's contribution color in depth order using $\alpha$-blending~\cite{max1995optical}. We refer to the computation of a single primitive's contribution at a single pixel, before $\alpha$-blending, as \textit{per-pixel evaluation}. For 3D Gaussians, per-pixel evaluation computes the primitive's \textit{response}, the unnormalized Gaussian density at a query location, and weights it by the primitive's opacity. The color of pixel $(x,y)$ is $c(x,y) = \sum_{i=1}^{N_t} \mathbf{c}_i \alpha_i T_i$, where $N_t$ is the number of 3D Gaussians in tile $t$'s list, $\mathbf{c}_i$ and $\alpha_i$ are the color and opacity of the $i$-th primitive, and $T_i = \prod_{j=1}^{i-1}(1 - \alpha_j)$ is the accumulated transmittance (the fraction of light remaining after previously blended primitives). The blend loop terminates when transmittance drops below a threshold, typically well before processing all $N_t$ primitives.

\begin{figure}[!htb] 
    \centering
    \includegraphics[width=\linewidth]{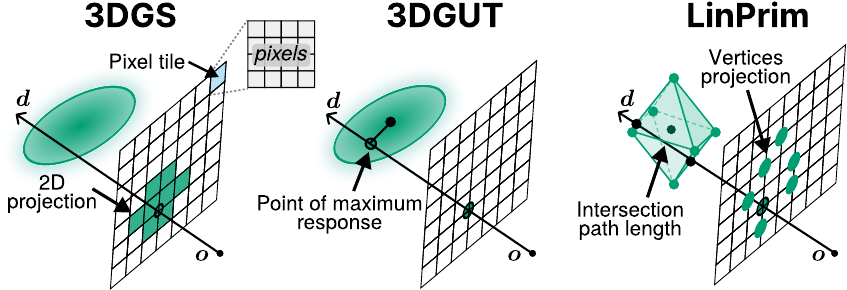}
    \caption{Projection and per-pixel evaluation for three representative point-based rendering methods.}
    \label{fig:background_fig_3}
\end{figure}

In Fig.~\ref{fig:background_fig_3}, we illustrate variations of point-based rendering methods using three representative examples: \textbf{3DGS} uses 3D Gaussian primitives. The projection of a 3D Gaussian onto the image plane produces a 2D elliptical footprint. Per-pixel evaluation computes the response of the projected 2D Gaussian at each pixel location. \textbf{3DGUT}~\cite{wu20253dgut} uses the same 3D Gaussian primitive but projects it by transforming a set of representative 3D points to estimate the 2D footprint. Per-pixel evaluation casts a ray with direction $\mathbf{d}$ from the camera origin $\mathbf{o}$ through each pixel and computes the maximum response of the Gaussian in 3D along the ray. \textbf{LinPrim}~\cite{von2025linprim} replaces the Gaussian with an octahedron. Projection maps the octahedron's vertices onto the image plane. Per-pixel evaluation similarly casts a ray but computes the contribution from the path length through the octahedron.

%% file: sections/Approach.tex
\section{\X Overview}
\label{sec:approach}

\subsection{Key Interfaces}
\label{sec:interface_to_the_programmer}

\X provides a programming interface that is designed for a programmer to  define \emph{how a single point-based primitive contributes to a single pixel's color}, without managing parallelization or hardware mapping. Thus, the simplest interface would be a single function \texttt{evaluate\_full}($p$, $px$, $cam$, $view$) that takes a primitive $p$, a pixel location $px=(x,y)$, camera parameters $cam$, and a camera \emph{view}, and returns the primitive's color contribution at that pixel. Rendering would then call \texttt{evaluate\_full} for every primitive–pixel pair and aggregate. This is prohibitively expensive: for $N$ primitives and $W\cdot H$ pixels {(image width $W$, height $H$)}, rendering requires $N \cdot W\cdot H$ calls (typically $>10^{12}$).

\X instead exploits two forms of redundancy that are present in every point-based rendering method. First, \texttt{evaluate\_full} is not jointly dependent on $p$ and $px$ everywhere: some sub-computations depend only on the primitive (and are therefore identical across all pixels), while others depend only on the pixel (and are identical across all primitives). \X separates these into two dedicated functions. \texttt{project}($p, cam, view$) processes each primitive once. \texttt{project} produces tile cull data $p_{tcd}$ and shader data $p_{sd}$. The shader data—typically a projected 2D position, 2D shape parameters, opacity, and view-dependent color—is reused across all pixels. \texttt{pixel\_info}($px, cam, view$) processes each pixel once and produces pixel data $px_{data}$ that is reused across all primitives. The per-primitive-per-pixel evaluation is then expressed as \texttt{evaluate}($px_{data}, p_{sd}$), which is called only on the subset of primitive–pixel pairs that actually interact. Second, each primitive affects only a small, spatially localized subset of pixels. \X exploits this by grouping pixels into rectangular \emph{tiles} and testing primitives against tiles rather than against individual pixels. A primitive that does not overlap with a tile is skipped for every pixel in that tile. The programmer expresses this tile-level test through \texttt{tile\_cull}($t_{\min}$, $t_{\max}$, $p_{tcd}$) that takes the tile boundaries $t_{\min}$ and $t_{\max}$ of tile $t$ and $p_{tcd}$ as input. \texttt{tile\_cull} returns whether primitive $p$ may contribute to any pixel in tile $t$.

\begin{figure}[!htb]
    \centering
    \includegraphics[width=\linewidth]{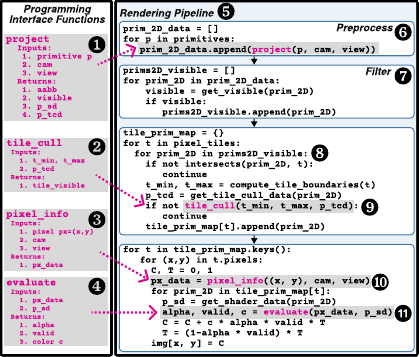}
    \caption{Overview of how the programmer interface functions exposed by \X are used in the rendering pipeline.}
    \label{fig:interface_overview}
\end{figure}

Together, the four functions—\texttt{project}, \texttt{tile\_cull}, \texttt{pixel\_info}, and \texttt{evaluate}—form \X's programming interface. The key design property is that each function operates on a single element: a single primitive, a single tile–primitive pair, or a single pixel. This means the programmer never reasons about parallelism, data layout, or scheduling, which is handled by \X automatically.
Fig.~\ref{fig:interface_overview} shows how the four functions are invoked within \X's four-stage pipeline~\circled{5}. In Preprocess~\circled{6}, \texttt{project}~\circled{1} runs once per primitive to produce $p_{tcd}$, $p_{sd}$, the depth $z$, the bounding box $aabb$, and a visibility flag $visible$. Filter~\circled{7} discards primitives that are not $visible$. The rasterizer builds, for each tile, a depth-sorted list of the primitives that may contribute to it: a framework-provided axis-aligned overlap test (intersect~\circled{8}) performs a conservative bounding-box check, and the user-defined \texttt{tile\_cull}~\circled{2}, \circled{9} optionally refines this at tile granularity. The shader runs \texttt{pixel\_info}~\circled{3}, \circled{10} once per pixel, then \texttt{evaluate}~\circled{4}, \circled{11} once per pixel–primitive pair over the gathered primitives in each pixel's tile, accumulating contributions into the final pixel color via $\alpha$-blending. 

The programming interface of \X decouples method-specific logic from the pipeline: swapping 3D Gaussians for octahedra, or switching from splatting-based to ray-based evaluation, requires only new implementations of the four functions; the pipeline, its parallelization, and its data structures remain unchanged. Because JAX provides automatic differentiation, no backward pass needs to be implemented—gradients flow through the four user-defined functions for free.
For 3DGS, the programmer instantiates only \texttt{project}, \texttt{tile\_cull}, and \texttt{evaluate}. \texttt{project} computes the 2D Gaussian parameters via EWA splatting~\cite{zwicker2002ewa}, and \texttt{evaluate} computes the per-pixel 2D Gaussian response (Listing~\ref{lst:xpr3dgs}). For LinPrim, \texttt{project} stores the \(L_1\)-slab parameters of the projected octahedron, \texttt{pixel\_info} is set to \texttt{None}, and \texttt{evaluate} reconstructs the pixel ray from the pixel coordinate using the \(+0.5\) pixel-center convention before returning the Beer-Lambert opacity from the ray's entry and exit depths through the octahedron. For 3DGUT, \texttt{project} applies the Unscented Transform to estimate the 2D footprint from seven sigma points; \texttt{pixel\_info} casts a ray from the camera origin through each pixel; and \texttt{evaluate} returns the analytic maximum Gaussian response along that ray (see~\secref{sec:listing_3dgut}). In all three cases, the remainder of the pipeline—filtering, rasterization, shading, and differentiation—is reused or unchanged.

\subsection{Exposing Parallelism in the Renderer to the Compiler}
\label{sec:exposing_parallelism}

\begin{figure}[!htb]
    \centering
    \includegraphics[width=0.95\linewidth]{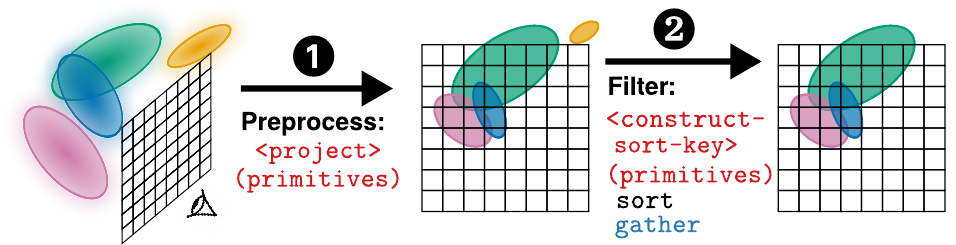}
    \caption{Parallelizing the preprocess and filter stages using \operations.}
    \label{fig:parallelization_preprocess}
\end{figure}

\X exposes hardware-agnostic parallelism in the rendering pipeline that can be used by a cross-platform compiler (XLA~\cite{openxla2024} in our implementation) to target a diverse set of compute hardware. To achieve this, \X decomposes the rendering pipeline into \emph{modules}. Every module falls into one of two categories. Modules whose computation can be parallelized across input data elements are expressed as \texttt{map\_fn(f)}. \texttt{map\_fn(f)} applies a scalar function \texttt{f} independently to each element of a batch, where an element may be a primitive, a pixel, a tile, a primitive–tile pair, or a primitive–pixel pair. Because invocations are independent, \texttt{map\_fn} exposes data parallelism that the compiler maps to SIMD execution on GPUs or to batched tensor operations on TPUs. A module can also comprise standard parallelization primitives: \gatherop, \scatterop, \prefixsumop, and \texttt{sort}. These primitives handle the steps that require inter-element communication, and optimized implementations are typically available across all target hardware.
A key additional constraint is that TPUs and similar accelerators require all tensor dimensions and loop trip counts to be known at compile time.  Several pipeline stages naturally produce data-dependent output sizes: the number of visible primitives after filtering and the total number of primitive–tile intersections both vary from frame to frame. \X handles this by fixing all such dimensions to compile-time upper bounds, padding unused entries, and replacing dynamic loops with fixed-trip-count loops that exit early. We now discuss how the rendering pipeline algorithm can be expressed as a composition of \operations, and how each module is parallelized using \texttt{map\_fn} or standard operations (\gatherop/\scatterop/ \prefixsumop/\texttt{sort}). We highlight each \texttt{map\_fn} scalar function in \textcolor[HTML]{D7191C}{\texttt{<red>}}, and indicate the elements it parallelizes over in braces (\textcolor[HTML]{D7191C}{\texttt{primitives}}, \textcolor[HTML]{D7191C}{\texttt{pixels}}, \textcolor[HTML]{D7191C}{\texttt{tiles}}, or \textcolor[HTML]{D7191C}{\texttt{intersections}}):

\noindent (1) \textbf{Preprocess.} \texttt{project} is applied independently to each of the $N$ primitives via a single \texttt{map\_fn}, as illustrated in Fig.~\ref{fig:parallelization_preprocess}. 

\noindent (2) \textbf{Filter.}  Starting from the per-primitive visibility flags computed in Preprocess, a \texttt{map\_fn} over the $N$ primitives builds a sort key from the primitive's flag and depth, a \texttt{sort} brings the visible primitives to the front in depth order, and a \gatherop collects them into a contiguous array of $M < N$ visible primitives.

\begin{figure}[!htb]
    \centering
    \includegraphics[width=\linewidth]{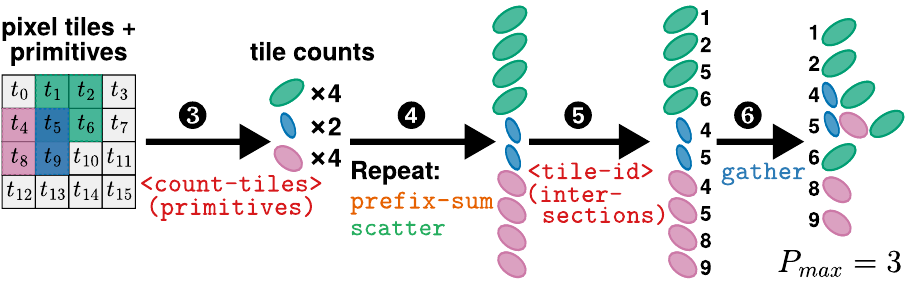}
    \caption{Modules used to parallelize the rasterizer.}
    \label{fig:parallelization_rasterizer}
\end{figure}

\noindent (3) \textbf{Rasterizer.} 
Between $T$ tiles of pixels and $M$ primitives, directly testing all $M \times T$ primitive–tile pairs for intersections is infeasible at scale ($M\sim 10^6$, $T\sim10^5$). Instead, \X exploits the fact that each primitive intersects only a small number of tiles—bounded by its bounding box area—and enumerates only those candidate intersections in a flat array of size $I$ (= number of tile intersections). A \texttt{map\_fn} over primitives counts tile intersections per primitive~\circled{3} (Fig.~\ref{fig:parallelization_rasterizer}). A \texttt{repeat} (\prefixsumop+\scatterop) \operation expands each primitive index by its count~\circled{4}; a second \texttt{map\_fn} assigns a unique tile ID to each intersection~\circled{5}; and a \gatherop assembles the per-tile primitive lists~\circled{6}. The full procedure, described in~\cref{sec:detailed_design_rasterizer}, produces a statically shaped $T\times P_{\max}$ matrix of primitive indices per tile.

\begin{figure}[!htb]
    \centering
    \includegraphics[width=0.9\linewidth]{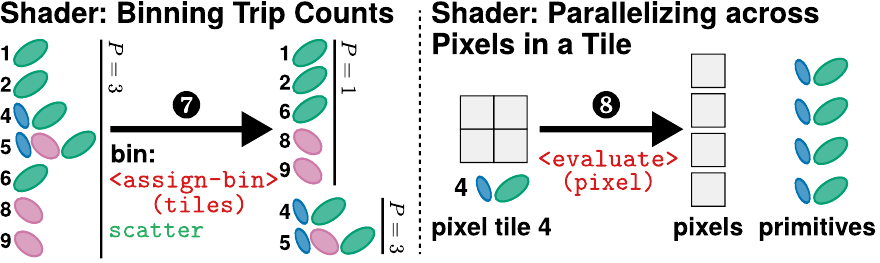}
    \caption{Parallelizing the shader using \operations.}
    \label{fig:parallelization_shader}
\end{figure}

\noindent (4) \textbf{Shader.}
All pixels within a tile iterate over the same primitive list, so the per-pixel color computation is parallelized via \texttt{map\_fn} over the pixels in a tile~\circled{8} ($2\times2$ pixel tile shown in Fig.~\ref{fig:parallelization_shader}). The remaining challenge is that different tiles intersect different numbers of primitives, so a single static loop bound applied uniformly would waste substantial computation on lightly loaded tiles. \X addresses this with \emph{trip-count binning}: tiles are sorted by primitive count and partitioned into multiple bins, each processed with a static trip count equal to its bin's upper bound~\circled{7}. Tiles within the same bin are parallelized via a second \texttt{map\_fn} over tiles. The full shader design is described in~\cref{sec:detailed_design_shader}.

%% file: sections/Detailed_Design.tex
\section{Detailed Design}
\label{sec:detailed_design}

\subsection{Preprocess and Filter}
\label{sec:detailed_design_perprimitive_operations}

\noindent\emph{Preprocess.} \X expresses this stage as \texttt{map\_fn} that uses \texttt{project} and $N$ primitives as arguments to expose parallelism. Each primitive is processed independently by hardware processing units. The operation \texttt{project} computes the projected footprint, bounding box, depth, a visibility flag, shader data, and tile cull data per primitive.

\noindent\emph{Filter.} The filter stage reduces the working set from $N$ total primitives to $M$ visible ones ($M < N$). Filter applies a \texttt{map\_fn} over $N$ primitives to extract the indices of visible primitives using their visibility flags. The \texttt{map\_fn} then uses these indices and the depths to construct a sort key that is used to sort the indices of visible primitives to the front in depth-sorted order. A gather then uses the sorted indices to collect the $M$ visible primitives into a contiguous array. The per-primitive tile counts of the $M$ visible primitives are passed to the rasterizer. The sum of these tile counts gives the total intersection count $I$, which the rasterizer uses to size its flat intersection array.

\subsection{Rasterizer}
\label{sec:detailed_design_rasterizer}

The rasterizer determines, for each pixel tile, which primitives can contribute to it. Because primitives are distributed unevenly across the image, the number of contributing primitives varies from tile to tile, and the natural representation is a per-tile list of variable length. Variable-length lists, however, require dynamic memory and dynamic control flow, neither of which is well supported on accelerators that compile to static tensor shapes. \X therefore reformulates rasterization as a sequence of \operations, producing a fixed-size matrix $Q$ of size $T \times P_{\max}$ in which row $t$ stores up to $P_{\max}$ depth-sorted primitive indices for tile $t$.

\begin{figure}[!h] 
    \centering
    \includegraphics[width=\columnwidth]{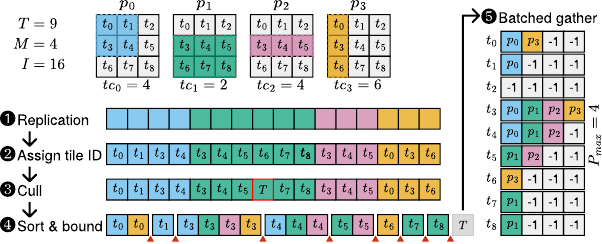}
    \caption{\X's rasterizer  using \operations over static-shape arrays.}
    \label{fig:fig_rasterizer}
\end{figure}

\X's rasterizer receives $M$ primitives $p_0, \ldots, p_{M-1}$ as input from the filter stage.
For each primitive $p_i$, \texttt{project} returns its depth $z_i$, its tile-space bounding box $[x_i^{\min}, x_i^{\max}) \times [y_i^{\min}, y_i^{\max})$, and its tile count $tc_i$ (the number of tiles its bounding box intersects). Fig.~\ref{fig:fig_rasterizer} shows the five processing stages of \X's rasterizer, \mbox{explained below}. 

\textbf{Replicate \ding{182}.}
\X prefix-sums the tile counts $tc_i$ to obtain each primitive's start offset $s_i$ in a flat array of size $I$. Each element of this flat array corresponds to a tile-primitive intersection. A \texttt{map\_fn} over the $I$ elements writes $p_i$ into slots $s_i, \dots, s_i + tc_i - 1$, so that $p_i$ appears exactly $tc_i$ times. For example, in Fig.~\ref{fig:fig_rasterizer}, primitive $p_0$ shown in blue is assigned to tiles $t_0, t_1, t_3$, and $t_4$.
\textbf{Assign tile ID \ding{183}.} We assign a unique tile ID for each intersection.
For a replicated entry with local offset $r$ within $p_i$'s bounding box, let $w_i = x_i^{\max} - x_i^{\min}$. XPR computes $x = x_i^{\min} + (r \bmod w_i)$ and $y = y_i^{\min} + \lfloor r / w_i \rfloor$, then assigns tile ID $t = yW' + x$, where $W'$ is the tile-grid width.
\textbf{Cull \ding{184}.}
XPR applies the optional \texttt{tile\_cull} function to each candidate primitive--tile pair, using the tile-cull data returned by \texttt{project}. Candidates that fail \texttt{tile\_cull} or the contribution threshold retain their slot but are reassigned tile ID $T$. If no \texttt{tile\_cull} function is provided, \X keeps all bounding-box candidates.
\textbf{Sort and identify bounds \ding{185}.}
\X lexicographically sorts the flat array by (tile ID, depth), placing rejected candidates at the end. For each valid tile $t_j$, the surviving candidates form a single contiguous segment whose primitive indices are ordered front-to-back. A \texttt{map\_fn} over the sorted tile IDs records each segment's start $b_j$ and count $c_j$. The visual gaps in Fig.~\ref{fig:fig_rasterizer} mark these segment bounds; they are not empty entries in the flat array.
\textbf{Gather \ding{186}.}
Using $b_j$ and $c_j$, XPR writes $Q[j, k] = P_{\text{sort}}[b_j + k]$ for $0 \le k < \min(c_j, P_{\max})$, where $P_{\text{sort}}$ is the sorted primitive-index array. XPR pads the remaining entries of $Q[j, :]$ with $-1$. The shader then reads $Q$, the counts $c_j$, and the per-primitive shader data, obtaining front-to-back primitive lists with bounded per-tile work and without $M \times T$ storage or dynamic per-tile lists.

\subsection{Shader}
\label{sec:detailed_design_shader}

The shader computes per-pixel colors by blending primitive contributions in depth order. It receives from the rasterizer (\secref{sec:detailed_design_rasterizer}) the $T \times P_{\max}$ depth-sorted primitive-index matrix $Q$, and for each pixel iterates over the primitives in its tile's row of $Q$, accumulating their contributions via $\alpha$-blending. Because the number of primitives per tile varies, the number of $\alpha$-blending iterations—the loop's \emph{trip count}—varies across tiles. Accelerators such as TPUs, however, require statically bounded loops, and using a single worst-case trip count across all tiles would waste substantial work on lightly loaded ones. XPR addresses this with \emph{trip-count binning}: tiles are grouped by primitive count, and each group is shaded with a single static trip count.

\begin{figure}[!htb]
    \centering
    \includegraphics[width=\linewidth]{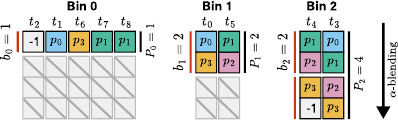}
    \caption{The shader receives $L$ bins. Each bin contains tiles whose primitive counts are bounded by one static trip count. The shader processes each tile's primitive list in batches of $b_l$ entries.}
    \label{fig:shader_detailedesign}
\end{figure}

\noindent\textit{Trip-count Binning.}\label{sec:trip-count-binning}
The idea behind trip-count binning is to replace a single global worst-case trip count with many tighter ones for different tiles. Thus, instead of using a single worst-case trip count $P_{\max}$, we define trip counts $P_0, P_1,\dots, P_{L-1}$. These $P_0, P_1,\dots , P_{L-1}$ are hyperparameters specified before rendering. \X assigns the $T$ tiles into $L$ bins according to their primitive counts: bin $l \in \{0, \ldots, L-1\}$ holds tiles with at most $P_l$ primitives, with $P_0 < P_1 < \dots < P_{L-1} = P_{\max}$. Tiles in the same bin are then shaded together with a single static trip count $P_l$, so lightly loaded tiles avoid the worst-case cost incurred in larger bins. For each bin, a gather collects the corresponding rows of $Q$ into a smaller fixed-size matrix with at most $P_l$ primitive entries per tile; entries beyond a tile's valid count are padded with sentinels and gated to zero opacity, so they do not affect the rendered color. Fig.~\ref{fig:shader_detailedesign} illustrates the scheme with $L = 3$ bins of sizes $P_0 = 1$, $P_1 = 2$, and $P_2 = 4$.
The batch size $b_l$ sets how many primitive entries the shader loads and processes at each step for bin $l$.
The shader exposes parallelism within each bin by mapping over tiles in the bin and, inside each tile, over its pixels. For each pixel $(x, y)$ in a tile of bin $l$, \X iterates through that tile's primitive list for up to $P_l$ iterations, the bin's static trip count.

Algorithm~\ref{alg:blend} describes the process for a bin $l$. For each tile $t$, the shader receives as input the trip count $P_l$, the batch size $b_l$, and the list $Q_t$ of primitive indices to iterate through. The shader loops across $Q_t$ primitives for pixel tile $t$ with a tile size of $w\times h$ (tile width $w$, height $h$) in bin $l$ using $\texttt{map\_fn}$ over pixels in the tile. The loop is structured as a two-level loop with a fixed outer trip count.

\begin{algorithm}[H]
\caption{Processing primitives for tile $t$ in bin $l$.}
\label{alg:blend}
\begin{algorithmic}[1]
\Require Trip count $P_l$, batch size $b_l$, shader data $sd$, list of primitive indices $Q_t$ for tile $t$, camera parameters $cam$, and camera $view$
\Ensure Rendered pixel colors for pixel tile $t$
\For{each pixel $px$ in tile $t$} 
    \Comment{parallelized via \texttt{map\_fn}}
  \State $px_{data} = \texttt{pixel\_info}(px, cam, view)$
  \For{$i = 0$ \textbf{to} $\lfloor P_l / b_l \rfloor -1$ }
      \Comment{$\lfloor P_l / b_l \rfloor$ iterations}
    \For{$j = 0$ \textbf{to} $b_l-1$} 
        \Comment{unrolled by factor $U$}
      \State $k= i\cdot b_l +j$
      \State $q = Q_t [k]$ \Comment{get primitive index}
      \State $p_{sd} = sd[q]$ 
         \Comment{fetch primitive shader data}
      \State $\alpha, valid, c \gets 
        \texttt{evaluate}(px_{data}, p_{sd})$
      \State $\ldots$ \Comment{gate and accumulate color}
    \EndFor
    \State $\ldots$ \Comment{check exit conditions at batch boundary}
  \EndFor
\EndFor
\end{algorithmic}
\end{algorithm}

%% file: sections/Methodology.tex
\section{Methodology}
\label{sec:methodology}

We implement \X in Python, using JAX~\cite{jax2018github} and the XLA~\cite{openxla2024} compilation framework. 

\noindent \textbf{Evaluation Platforms and Datasets.} We evaluate \X on the NVIDIA L40S GPU (\texttt{L40}), AMD MI210 GPU (\texttt{MI210}), Intel Max 1100 GPU (\texttt{MAX1100}), and Google v6e TPU (\texttt{TPU}). 
We present an evaluation of nine real-world scenes from the Mip-NeRF360~\cite{barron2022mip} and Tanks\&Temples~\cite{knapitsch2017tanks} datasets, and eight synthetic scenes from the Blender~\cite{mildenhall2021nerf} dataset.

\noindent\textbf{Baselines, Configurations, and Metrics.}
We evaluate three \X implementations against the corresponding authors' released CUDA codebases: \texttt{\X-3DGS} against 3DGS~\cite{kerbl20233d,3dgsbase_code}, \texttt{\X-3DGUT} against 3DGUT~\cite{wu20253dgut,3dgutbase_code}, and \texttt{\X-LinPrim} against LinPrim~\cite{von2025linprim,linprimbase_code}. We refer to these codebases as \texttt{Base}. We train with \texttt{Base} using the authors' standard configuration, then render the same trained model with both \texttt{Base} and \X. We report PSNR/SSIM/LPIPS and FPS on test views; the relative speed of \X compared to \texttt{Base} is reported as the mean of per-scene FPS ratios. Each \texttt{Base} renderer is a heavily optimized CUDA implementation specialized for one rendering method, with hand-written forward and backward kernels; together, these baselines provide a strong comparison for \X, which expresses all three methods in a single framework. We evaluate two configurations of \X with different maximum intersections per tile, $P_{\max}\in\{4096,8192\}$ (see (3) Rasterizer in~\cref{sec:exposing_parallelism}), denoted \texttt{\X-4K} and \texttt{\X-8K}.

\noindent\textbf{Hyperparameters.}\label{sec:hyperpara_selection} Expressing the rendering pipeline as static-shape operations is what enables \X to execute on accelerators that lack dynamic control flow, such as TPUs. This design requires that all tensor dimensions and loop trip counts are fixed at compile time. Two parameters in particular are data-dependent and vary across views: the maximum number of visible primitives $M$ and the maximum sum of per-primitive tile counts $I$. Four additional hyperparameters affect rendering speed: tile size $w \times h$, unroll factor $U$, number of bins $L$, and batch divisor $D$. The tile size $w \times h$ specifies the shape of each pixel tile, with width $w$, height $h$, and $wh$ pixels per tile. The number of bins $L$ determines how many shader trip-count bins are used to group tiles by primitive count. For bin $l$, the batch divisor $D$ is used to derive the shader batch size $b_l$ from the bin trip count $P_l$, with $b_l = \lfloor P_l/D \rfloor$ rounded to a multiple of $U$. The unroll factor $U$ specifies the XLA unroll factor for the inner loop over the $b_l$ primitive entries in each shader batch. \X's hyperparameter selection determines all six settings automatically, so the user does not need to configure any hyperparameters manually. The hyperparameter selection takes as input the set of views to be rendered and the trained primitive parameters, and tunes all six necessary settings that are reused for all subsequent processing of that scene on that hardware.

Table~\ref{tab:hyperparameters} summarizes the \X hyperparameter search space. \(P_{\max}\) is a global accuracy/speed tuning knob shared by all three methods. $L$, $D$, and $U$ are selected per scene by the hyperparameter selection, enabling the configuration to adapt to scene geometry without manual tuning.

\begin{table}[!htb] 
    \centering
    \small
    \setlength{\tabcolsep}{4pt}
    \begin{tabular}{l|c}
        \textbf{Hyperparameter} & \textbf{Value / Search Space} \\
        \hline
        Tile size (GPU)                           & \{$8\times 8,$\,$16\times 16$\} \\
        Tile size (TPU)                          & \{$16\times 16,$\,$32\times 32$\} \\
        \(P_{\max}\)                             & $\{4096,\ 8192\}$ \\
        Number of bins ($L$)                & $\{2,4,6,8\}$ \\
        Batch divisor ($D$)                 & $\{1, 2,4,8\}$ \\
        XLA unroll factor ($U$)             & $\{1,2,4,8,16,32\}$ \\
        Bin upper bounds (for GPU)          & $64,128,256,512,768,1024,2048$ \\
        Bin upper bounds (for TPU)          & $256,512,768,1024,1280,1536,2048$ \\
    \end{tabular}
    \caption{\X\ hyperparameter search space. $L$, $D$, $U$ are selected per scene by the hyperparameter selection; batch sizes are derived as $\mathrm{b}_l=\lfloor P_l/D\rfloor$ rounded to a multiple of $U$. \(P_{\max}\) is the tuning knob (accuracy/speed) and is identical for all methods.}
    \label{tab:hyperparameters}
\end{table}

The hyperparameter selection operates in two steps. First, it renders every target view once (forward-only, no gradients) and records $M$, $I$, and the number of primitives each tile intersects for each view. It sets $M$ and $I$ to the observed maxima across all views. Profiling all views is necessary because any view whose values exceed the chosen bound can produce incorrect rendered images or a drop in accuracy. Since the hyperparameter selection profiles the same views it will render, the bounds are guaranteed to be sufficient without any safety margin or overestimation heuristic. Second, the hyperparameter selection searches the four performance parameters in five sequential phases: the tile size (two candidates; each needs to be profiled since different tile sizes change the tile grid), the unroll factor $U$ (six candidates, benchmarked exhaustively), the number of bins $L$ (four candidates, benchmarked exhaustively, with a 2\% noise-tolerance tie-breaker that re-benchmarks the lowest and highest tied candidates with additional runs), a search of $U$ at the chosen $L$ (six candidates, triggered whenever $L$ shifts the bin layout, since a larger unrolled scan body shifts the tradeoff between instruction-level parallelism and register pressure), and the batch divisor $D$ (up to four candidates, preferring $D{=}4$ within noise tolerance). The per-tile counts from profiling also determine the data-aware assignment of tiles to bins, computed from worst-case cumulative counts across all views. In total, the full hyperparameter selection (profiling and search) takes only a few minutes, and this is a one-time cost that is negligible relative to the training time of the scene.

%% file: sections/Evaluation.tex
\section{Evaluation}
\label{sec:evaluation}

\subsection{Ease of Implementing Rendering Methods}
\label{sec:ease_of_implementing_rendering_methods}

We demonstrate that \X can express various point-based rendering methods with only small code changes. \texttt{\X-3DGS} requires implementing \texttt{project}, \texttt{evaluate}, and \texttt{tile\_cull}. The \texttt{project} function computes the 2D Gaussian projection parameters via EWA splatting, and \texttt{evaluate} computes the per-pixel 2D Gaussian response. All other framework modules are reused unchanged. We demonstrate how \texttt{XPR-3DGUT} and \texttt{XPR-LinPrim} are implemented in Appendix~\ref{sec:ease_of_implementing_rendering_methods_appendix}.

\input{listings/3dgs}

Table~\ref{tab:lines_of_code} reports the lines of code required to implement these methods in \X compared to \texttt{Base}. We count non-blank lines, excluding comments, docstrings, and import/\#include directives. For \X, we count lines across programmer interface functions and any helper functions they call. \texttt{Base} (non-shared vs 3DGS) shows the number of new lines of code added over the 3DGS codebase. Each \X implementation requires only 260 to 304 lines of Python, significantly fewer than \texttt{Base}.

\input{tables/loc}

\subsection{Quality and Performance of \X}
\label{sec:quality_performance}

To verify that XPR's implementations are equivalent to the \texttt{Base} implementations, we evaluate rendering quality on test views of the scenes from the three datasets in terms of PSNR/SSIM/LPIPS and rendering speed in terms of FPS averaged across all scenes in a dataset on an \texttt{L40} (Table~\ref{tab:quality_comparison}). We report SSIM/LPIPS measurements in Appendix~\ref{sec:quality_appendix} since \X matches \texttt{Base}.

\input{tables/quality_comparison}

First, we observe that \texttt{\X-8K} matches \texttt{Base-3DGS} and \texttt{Base-LinPrim} on all three datasets in PSNR, and stays within $0.06$ PSNR compared to \texttt{Base-3DGUT}, which we attribute to numerical noise. Second, lowering $P_{\max}$ to 4096 leads to higher frame rates in most configurations because this reduces the shader's $\alpha$-blending iterations. \texttt{\X-4K} (the default setting) averages $1.08\times$ the frame rate of \texttt{\X-8K} (up to $1.38\times$). A small quality loss, however, is incurred since some primitives' contributions may be skipped: \texttt{\X-4K} remains within $0.06$ PSNR of \texttt{Base} on \texttt{360} and $0.12$ PSNR on \texttt{T\&T}, and within $0.58$ PSNR on \texttt{Synthetic}, where the larger gap is attributed to LinPrim's ray-intersection evaluation being more sensitive. Third, the frame rates for \texttt{XPR-3DGS} in comparison to \texttt{Base} are $0.95\times$ on \texttt{360}, $0.96\times$ on \texttt{T\&T}, and $0.76\times$ on \texttt{Synthetic} (for $P_{\max}=4096$). JAX incurs a small kernel launch overhead that is more significant when frame rates are very high, as in \texttt{Synthetic}. \X's overheads are higher on \texttt{3DGUT} ($69.52$\% of \texttt{Base} on average) because \texttt{3DGUT}'s pixel evaluation depends on per-pixel world-space ray directions, implemented via \texttt{pixel\_info} in \X (Listing~\ref{lst:xpr3dgut}). \texttt{Base} fuses this per-pixel computation into blending, making it faster. \texttt{3DGS} avoids this cost because its evaluation depends only on 2D pixel coordinates.
\texttt{\X-LinPrim} outperforms \texttt{Base} because we implement a more aggressive \texttt{tile\_cull} strategy that prunes more primitive–tile intersections, reducing computation (Listing~\ref{lst:xprlinprim}). This demonstrates that \X's high-level interface enables rapid prototyping of efficient rendering strategies.

Thus, we demonstrate that \X enables prototyping and exploration of a wide range of point-based differentiable rendering methods using high-level abstractions and fewer lines of code, while achieving per-scene frame rates that are, on average, within $2.82\%$ of those of hand-written CUDA implementations, and up to \mbox{$208.18\%$ faster}.

\noindent \textbf{Demonstrating \X's portability across platforms.}
\label{sec:results_platforms}
Table~\ref{tab:platform_fps_by_tflops} reports the average FPS of \texttt{\X-4K}, directly compiled from the same Python codebase via XLA without any platform-specific modifications, for all three datasets on four hardware targets. To our knowledge, \X is the first point-based differentiable renderer that can execute on all four platforms and provides the only available implementation on Intel data center GPUs and Google TPUs. Absolute frame rates vary across platforms, reflecting two factors: the maturity of XLA's backend for each target and the available FP32 compute throughput. On the TPU, single-chip performance (Table~\ref{tab:platform_fps_by_tflops}) is lower than on GPUs, but each stage of XPR's pipeline is parallelizable across chips to scale performance in \mbox{multi-chip systems}.

\input{tables/XPR_backend_comparison}

%% file: listings/3dgs.tex
\begin{lstlisting}[style=python, caption={Implementation of \texttt{\X-3DGS}.}, label=lst:xpr3dgs]
def project(p, cam, view, cfg):
  # Get world to camera projection matrix
  R_w2c = world_2_camera_projection_matrix(view)
  mu_c = world_2_cam(p.mu, R_w2c)
  mu_2D  = compute_mu_2D(mu_c, cam)
  # EWA splatting 
  cov_2D, det, _ = compute_cov_2D(p.s, p.q, mu_c, R_w2c, cam, cfg)
  con_2D = compute_conic_2D(cov_2D, det)
  
  # Compute extent, radius, and bounding box
  ex, ey = compute_extents(cov_2D, det, p.o, cfg)
  r = compute_radius(ex, ey)
  aabb = get_bounding_box(mu_2D, ex, ey, cfg)
  
  # Compute tile count
  tc = (aabb.x_max-aabb.x_min) * (aabb.y_max-aabb.y_min)
  
  # Compute visibility flag and adjust tile count
  visible = (det>0) & (p.o>=cfg.alpha_thresh) & (r > 0) & (tc>0) & (mu_c[2]>cfg.z_near)
  tc *= float(visible)

  # Compute view-dependent color c
  c = compute_color(p.mu, p.sh, view)
  # Produce tile cull data and shader data for primitive p
  p_tcd = concat(con_2D[0,0], con_2D[0,1], con_2D[1,1], p.o, mu_2D, visible)
  p_sd = concat(mu_2D[0], mu_2D[1], -0.5*con_2D[0,0], -con_2D[0,1], -0.5*con_2D[1,1], p.o, c)
  
  return ProjectResult(depth=mu_c[2], visible=visible, tile_cull_data=p_tcd, shader_data=p_sd, aabb=aabb, tile_count=tc)

def tile_cull(tile_min, tile_max, p_tcd, cfg):
  conic, o, mu_2D, visible = unpack(p_tcd)
  mu_outside = (mu_2D[0] < tile_min[0]) | (mu_2D[0] > tile_max[0]) | (mu_2D[1] < tile_min[1]) | (mu_2D[1] > tile_max[1])
  
  # Compute location of the 2D point inside the tile that maximizes the response
  point_2D = compute_point_with_max_response(tile_min, tile_max, conic, mu_outside)

  # Compute max contribution power of 2D Gaussian 
  dx = mu_2D[0]-point_2D[0]
  dy = mu_2D[1]-point_2D[1]
  power = 0.5*(dx*dx*conic[0] + 2*dx*dy*conic[1] + dy*dy*conic[2])
  
  # Keep tile iff max contribution to tile satisfies: o * exp(-power) >= cfg.alpha_thresh
  tile_visible = (power <= log(o / cfg.alpha_thresh)) & visible
  return tile_visible
  
def evaluate(px_data, p_sd):
  mu_2D, conic, o, c = unpack(p_sd)
  dx = px_data[0] - mu_2D[0]
  dy = px_data[1] - mu_2D[1]
  power =  dx * dx * conic[0] + dx * dy * conic[1] + dy * dy * conic[2]
  alpha = o * exp(power)
  valid = power<=0.
  return EvaluateResult(alpha=alpha, valid=valid, color=c)

XPR_3DGS = MethodSpec((*@project@*)=project, (*@tile\_cull@*)=tile_cull, (*@pixel\_info@*)=None, (*@evaluate@*)=evaluate)
\end{lstlisting}

%% file: tables/loc.tex
\begin{table}[!htb]
      \centering
      \scriptsize
      \begin{tabular}{lrrr}
          \toprule
           & \texttt{3DGS} & \texttt{LinPrim} & \texttt{3DGUT} \\
          \midrule
          \X (Python)                  &   260 &   304 &   297 \\
          \texttt{Base} (total)                 & 2{,}153 & 2{,}516 & 4{,}915 \\
          \texttt{Base}  (non-shared vs 3DGS)  &   --- &   862 & 4{,}122 \\
          \bottomrule
      \end{tabular}
      \caption{Lines of code comparison of XPR with the corresponding \texttt{Base} implementations for different point-based rendering methods.}
      \label{tab:lines_of_code}
  \end{table}

%% file: tables/quality_comparison.tex
\begin{table}[!htb]
    \centering
    \setlength{\tabcolsep}{4pt}
    \resizebox{\columnwidth}{!}{
    \begin{tabular}{llccccccccc}
        \toprule
         & Dataset & \multicolumn{3}{c}{\texttt{360}} & \multicolumn{3}{c}{\texttt{T\&T}} & \multicolumn{3}{c}{\texttt{Synthetic}} \\
        \cmidrule(lr){3-5} \cmidrule(lr){6-8} \cmidrule(lr){9-11}
        Method & Metric & \texttt{Base} & \texttt{\X-4K} & \texttt{\X-8K}
                        & \texttt{Base} & \texttt{\X-4K} & \texttt{\X-8K}
                        & \texttt{Base} & \texttt{\X-4K} & \texttt{\X-8K} \\
        \midrule
        \multirow{2}{*}{\texttt{3DGS}}
            & PSNR  & 29.11  & 29.11  & 29.11  & 23.71  & 23.64  & 23.71  & 33.45  & 33.41  & 33.45  \\
            & FPS   & 100.81 & 95.16  & 92.02  & 130.16 & 124.59 & 106.76 & 311.71 & 232.68 & 213.96 \\
        \midrule
        \multirow{2}{*}{\texttt{3DGUT}}
            & PSNR  & 28.92  & 28.86  & 28.86  & 23.77  & 23.65  & 23.72  & 33.47  & 33.47  & 33.47  \\
            & FPS   & 105.66 & 71.55  & 69.69  & 97.74  & 78.37  & 68.41  & 349.80 & 190.28 & 179.51 \\
        \midrule
        \multirow{2}{*}{\texttt{LinPrim}}
            & PSNR  & 27.94  & 27.93  & 27.94  & 22.22  & 22.22  & 22.22  & 33.20  & 32.62  & 33.20  \\
            & FPS   & 66.31  & 71.39  & 69.73  & 89.06  & 99.17  & 102.09 & 73.10  & 116.42 & 99.28  \\
        \bottomrule
    \end{tabular}
    }
    \caption{Comparison of average PSNR and FPS of \X with \texttt{Base} implementations for different point-based rendering methods and datasets on an \texttt{L40}.}
    \label{tab:quality_comparison}
\end{table}

%% file: tables/XPR_backend_comparison.tex
\begin{table}[H]
    \centering
    \scriptsize
    \setlength{\tabcolsep}{4pt}
    \resizebox{\columnwidth}{!}{%
    \begin{tabular}{l ccc ccc ccc}
        \toprule
        & \multicolumn{3}{c}{\texttt{3DGS}} & \multicolumn{3}{c}{\texttt{3DGUT}} & \multicolumn{3}{c}{\texttt{LinPrim}} \\
        \cmidrule(lr){2-4} \cmidrule(lr){5-7} \cmidrule(lr){8-10}
        \textbf{Platform} & \texttt{360} & \texttt{T\&T} & \texttt{Syn.} & \texttt{360} & \texttt{T\&T} & \texttt{Syn.} & \texttt{360} & \texttt{T\&T} & \texttt{Syn.} \\
        \midrule
        \texttt{MI210}   & 29.60 & 40.67  & 67.16  & 23.71 & 33.98 & 73.14  & 22.91  & 34.39 & 33.79 \\
        \texttt{L40}     & 95.16 & 124.59 & 232.68 & 71.55 & 78.37 & 190.28 & 71.39 & 99.17 & 116.42 \\
        \texttt{MAX1100} & 9.95  & 12.83  & 18.59  & 6.07  & 6.92  & 11.00  & 9.36  & 15.33  & 16.68  \\
        \texttt{TPU}     & 1.89 & 2.22  & 6.19  & 1.52 & 1.19 & 6.92  & 1.70 & 3.00 & 3.02 \\
        \bottomrule
    \end{tabular}%
    }
    \caption{Average rendering FPS of \X-4K on various platforms.}
    \label{tab:platform_fps_by_tflops}
\end{table}

%% file: sections/Discussion.tex
\section{Discussion}
\label{sec:discussion}

The primary goal of \X is to enable rapid prototyping, exploration, and code reproduction across a wide range of platforms. \X is designed for compatibility with existing cross-platform compilers and ML IRs. Thus, compiler optimizations and hardware-specific performance tuning are outside of the scope of this work. A key resulting limitation is that rendering efficiency on any given platform is bounded by the maturity of the compiler backend for that target. \X currently uses JAX/XLA~\cite{openxla2024} with HLO as the intermediate representation, and supports any hardware target with an XLA backend, including NVIDIA, AMD, and Intel GPUs, TPUs, Trainium, and CPUs (we present evaluation on a representative subset). However, any compiler stack that can efficiently lower \X's map\_fn, scatter, gather, and sort operations is a viable backend—for example, TorchDynamo with ATen IR~\cite{ansel2024pytorch}. In terms of generality and expressiveness, \X is powerful enough to express a wide range of point-based rendering techniques (examples non-comprehensively listed in Table~\ref{tab:x-generality}). 
Techniques that involve ray marching or ray tracing (e.g., NeRFs, physically-based rendering) cannot be expressed with \X.    

%% file: sections/Conclusion.tex
\section{Conclusion}
\label{sec:conclusion}

We presented \X, an extensible point-based differentiable rendering framework that separates method-specific logic from the shared rendering pipeline. \X lowers the barrier to exploring new point-based rendering techniques, learning/teaching differentiable rendering, and using a wider range of hardware platforms for differentiable rendering. Integration with existing ML compiler stacks also makes differentiable rendering easier to incorporate as a component within larger optimization and learning systems.

%% file: sections/Appendix.tex
\section{Evaluation Details}

\subsection{Ease of Implementing Rendering Methods}
\label{sec:ease_of_implementing_rendering_methods_appendix}

We demonstrate that \X can express various point-based rendering methods with only small code changes. \X's high-level programming interface separates method-specific logic from the shared rendering pipeline. \X exposes two main data structures: (1) The \texttt{PrimitiveParams} data structure contains data fields that define the shape and density of the primitive. \texttt{PrimitiveParams} consists of required attributes for a primitive such as a 3D position $\mu$, opacity $o$,  and spherical harmonic coefficients $sh$. The structure can be extended with custom data fields. (2) The \texttt{MethodSpec} data structure consists of the four functions—\texttt{project}, \texttt{tile\_cull}, \texttt{pixel\_info}, and \texttt{evaluate}—that form \X's programming interface. \texttt{MethodSpec} specifies the rendering method, i.e., \emph{how} the primitive is processed and rendered. Thus, \X allows the programmer to build a new rendering method by providing the data fields in \texttt{PrimitiveParams} and implementing the four functions that are used to initialize the rendering method with \texttt{MethodSpec}.

\input{sections/Implementation_XPR3DGS}
\input{sections/Implementation_XPR3DGUT}

\input{sections/Implementation_XPRLinPrim}

\subsection{Quality of XPR}
\label{sec:quality_appendix}

 \input{tables/quality_comparison_appendix}

\subsection{Ablation Study}
\label{sec:ablation}

\begin{figure}[!htb]
    \centering
    \includegraphics[width=\linewidth]{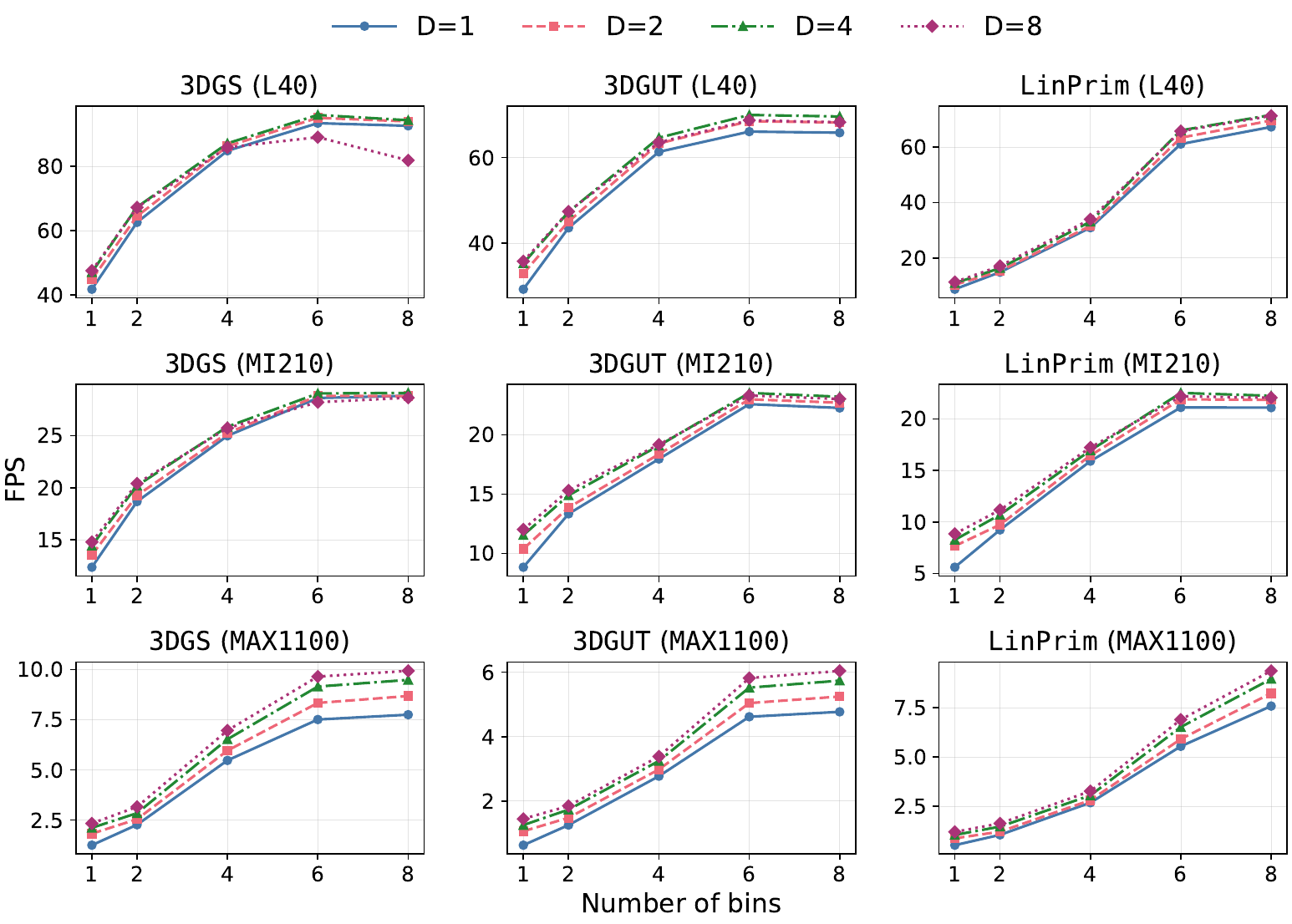}
    \caption{Average rendering FPS (y-axis) vs. the number of bins (x-axis) with different batch divisors for the \texttt{360} dataset.}
    \label{fig:bins_vs_performance}
\end{figure}

\noindent We study how different hyperparameters of \X affect performance across various devices. Fig.~\ref{fig:bins_vs_performance} depicts the average FPS of \X across different rendering methods and devices (\texttt{L40}, \texttt{MI210}, \texttt{MAX1100}) for the \texttt{360} dataset. First, we observe that increasing the number of bins increases FPS initially. This is because, with fewer bins, computing pixel colors requires more redundant computation beyond the actual number of primitives that the tiles in each bin intersect. For some pixel tiles, the blending loop must process more primitives without the ability to terminate early. As the number of bins increases, the FPS plateaus or even drops. Beyond the optimal bin count, shrinking per-bin batch sizes reduces SM occupancy and increases launch overheads faster than the additional bins reduce redundancy. Second, the batch divisor controls the granularity of early exit in the blend loop. A smaller batch divisor checks exit conditions less frequently, reducing per-iteration overhead at the cost of wasted work on already-saturated pixel colors. A larger batch divisor enables tighter early termination but adds loop overhead. Since both the number of bins and the batch divisor affect performance across scenes and hardware, \X's hyperparameter selection process sweeps these hyperparameters at compile time and chooses a performant configuration for a given scene and device.

%% file: sections/Implementation_XPR3DGS.tex
\subsubsection{Implementation of \X-3DGS}
\label{sec:listing_3dgs}
\noindent  Each \textbf{3DGS}\label{sec:3dgs}~\cite{kerbl20233d}  primitive is a 3D Gaussian density with center $\mu$, covariance $\Sigma=RSS^TR^T$ (where $R$ is a rotation matrix and $S$ a diagonal scaling matrix), opacity $o$, and spherical harmonic coefficients $sh$ for view-dependent color. In Listing~\ref{lst:xpr3dgs_params}, we show the data structure \texttt{PrimitiveParams} of \texttt{\X-3DGS}.

\begin{lstlisting}[style=python, basicstyle=\scriptsize\ttfamily, caption={\texttt{\X-3DGS}: \texttt{PrimitiveParams}.}, label=lst:xpr3dgs_params]
class PrimitiveParams:
  mu : Array  # (3,)  mean in world space
  s  : Array  # (3,)  scales
  q  : Array  # (4,)  unit quaternion for rotation
  sh : Array  # (C,3) spherical harmonic coefficients 
  o  : Array  # (1,)  opacity
\end{lstlisting}

\texttt{\X-3DGS} implements \texttt{project}, \texttt{evaluate}, and \texttt{tile\_cull} (see Listing~\ref{lst:xpr3dgs}). \texttt{\X-3DGS} sets \texttt{pixel\_info}{=}\texttt{None}, since per-pixel evaluation depends only on the 2D pixel coordinates. The \texttt{project} function computes the projection of a 3D Gaussian onto the image plane via EWA splatting \cite{zwicker2002ewa}, yielding a 2D Gaussian density with mean $\mu_{2D}$ and covariance $\Sigma_{2D}$ given by $G_{2D}(\mathbf{x}) = \exp\bigl(-\frac{1}{2}(\mathbf{x}-\mu_{2D})^\top \Sigma_{2D}^{-1} (\mathbf{x}-\mu_{2D})\bigr)$ where $\Sigma_{2D}^{-1}$ is the inverse of the projected covariance, referred to as the \emph{conic matrix}. The \texttt{project} function then computes the bounding box $aabb$, the tile count $tc$, a visibility flag $visible$, and produces the tile cull data $p_{tcd}$ and shader data $p_{sd}$ for primitive $p$. The implementation of the \texttt{tile\_cull} function follows the approach given by StopThePop \cite{radl2024stopthepop}: for each tile in the 3D Gaussian's bounding box $aabb$, \texttt{tile\_cull} computes the 2D location $\mathbf{x}$ inside the tile that maximizes the contribution $\alpha=o\cdot G_{2D}(\mathbf{x})$. Since the 2D Gaussian's response falls off smoothly with increasing distance to $\mu_{2D}$, the 2D location that maximizes the contribution is either $\mu_{2D}$ (if $\mu_{2D}$ is located inside the tile) or a point along the edges of the tile facing $\mu_{2D}$. The tile-primitive pair is culled if the contribution $\alpha$ falls below a threshold $\alpha_{\mathrm{thresh}}=\frac{1}{255}$. The \texttt{evaluate} function computes the contribution $\alpha=o\cdot G_{2D}(px)$ at a pixel query location $px=(x,y)$ and a validity flag $valid$. All other framework modules are reused unchanged. 

%% file: sections/Implementation_XPR3DGUT.tex
\subsubsection{Implementation of \X-3DGUT}
\label{sec:listing_3dgut}

\noindent \textbf{3DGUT}\label{sec:3dgut}~\cite{wu20253dgut} uses the same 3D Gaussian primitive as 3DGS but changes how primitives are projected and evaluated. Instead of linearizing the projection via the Jacobian as in EWA splatting, 3DGUT applies the Unscented Transform (UT)~\cite{gustafsson2011some}: it selects sigma points from the 3D Gaussian, transforms each through the camera projection, and estimates the projected mean $\mu_{2D}$ and covariance $\Sigma_{2D}$ in pixel space from the weighted projected points. The resulting 2D Gaussian is used for tile assignment and tile culling. For per-pixel evaluation, 3DGUT computes the Gaussian response in 3D along a camera ray $\mathbf{r}(\tau)=\mathbf{o}+\tau\mathbf{d}$, where $\mathbf{o}$ is the origin of the camera and $\mathbf{d}$ is the direction of the ray. The depth that maximizes the response is
\begin{equation}
    \tau_{\max} = \frac{(\mu-\mathbf{o})^\top \Sigma^{-1} \mathbf{d}}{\mathbf{d}^\top \Sigma^{-1} \mathbf{d}} = \frac{-\mathbf{o}_\text{c}^\top \mathbf{d}_\text{c}}{\mathbf{d}_\text{c}^\top \mathbf{d}_\text{c}},
\end{equation}
where $\mathbf{o}_\text{c}=S^{-1}R^{\top}(\mathbf{o}-\mu)$ and $\mathbf{d}_\text{c}=S^{-1}R^{\top}\mathbf{d}$ are the ray origin and direction in the canonical space of the Gaussian~\cite{wu20253dgut,kerbl20233d}. \texttt{XPR-3DGUT} implements all four functions of \X's programming interface (see Listing~\ref{lst:xpr3dgut}). The \texttt{project} function computes the UT footprint and stores the projected conic, opacity adjusted following Mip-Splatting~\cite{yu2024mip}, and mean in tile cull data $p_{tcd}$ for primitive $p$; it also precomputes $\mathbf{o}_\text{c}$ and $M=S^{-1}R^\top$ in shader data $p_{sd}$. The \texttt{tile\_cull} function follows \texttt{XPR-3DGS} after applying the ray-offset adjustment to the tile bounds. The \texttt{pixel\_info} function computes the ray direction $\mathbf{d}$ at a pixel, and \texttt{evaluate} uses the precomputed $p_{sd}$ to compute $\mathbf{d}_\text{c}=M\mathbf{d}$ and the contribution $\alpha=o \cdot G(\mathbf{o}+\tau_{\max}\mathbf{d})$. All other framework modules are reused unchanged.

\input{listings/3dgut}

%% file: listings/3dgut.tex
\begin{lstlisting}[style=python, basicstyle=\scriptsize\ttfamily, caption={Implementation of \texttt{\X-3DGUT}.}, label=lst:xpr3dgut]
def project(p, cam, view, cfg):
 # Get world to camera projection matrix
  R_w2c = world_2_camera_projection_matrix(view)
  # Construct rotation matrix R
  R = get_R(p.q)

  ## Unscented Transform
  # Compute sigma points in world space
  sigma_points = compute_sigma_points(p.mu, p.s, R, cfg)
  # Project sigma points to camera space
  sigma_points_c = vectorize(world_2_cam)(sigma_points, R_w2c)
  mu_c = sigma_points_c[0]
  # Project to pixel space
  sigma_points_2D = vectorize(compute_mu_2D)(sigma_points_c, cam)
  # Estimate mu_2D and cov_2D to approximate 2D Gaussian
  mu_2D, cov_2D = approximate_2D_gaussian(sigma_points_2D, cfg)
  det_orig = determinant(cov_2D)
  # Apply low-pass filter to cov_2D (antialiasing) 
  cov_2D = lowpass_filter(cov_2D)
  det = determinant(cov_2D)
  con_2D = compute_conic_2D(cov_2D, det)
  
  # Adjust opacity (Mip-Splatting)
  o = p.o * mip_splatting_2D(det_orig, det)
  
  # Compute extent, radius, and bounding box
  ex, ey = compute_extents(cov_2D, det, o, cfg)
  r = compute_radius(ex, ey)
  # Account for ray offset
  aabb = get_bounding_box(mu_2D - 0.5, ex, ey, cfg) 

  # Compute tile count
  tc = (aabb.x_max-aabb.x_min) * (aabb.y_max-aabb.y_min)

  # Compute view-dependent color c
  c = compute_color(p.mu, p.sh, view)
  
  # Compute visibility flag and adjust tile count
  visible = (det>0) & (o>cfg.alpha_thresh) & (r>0) & (tc>0) & (mu_c[2]>cfg.z_near)
  tc *= float(visible)

  # Produce tile_cull_data and shader_data for primitive p
  p_tcd = concat(con_2D[0,0], con_2D[0,1], con_2D[1,1], o, mu_2D, visible) 
  M = diag(1.0/p.s) @ R.T
  cp = -R_w2c[:3,:3].T @ R_w2c[:3,3]
  ro_canonical = (M @ (cp - p.mu)).flatten() 
  p_sd = concat(ro_canonical, M.flatten(), o, c)

  return ProjectResult(depth=mu_c[2], visible=visible, tile_cull_data=p_tcd, shader_data=p_sd, aabb=aabb, tile_count=tc)

def tile_cull(tile_min, tile_max, p_tcd, cfg):
  # Adjust tile bounds for ray offset
  tile_max[0] += 1 # tile_max_x
  tile_max[1] += 1 # tile_max_y
  # Rest of tile_cull in XPR-3DGS can be reused unchanged
  ...

def pixel_info(px, cam, view, cfg):
  # Camera ray direction
  dc = [(px[0] + 0.5 - cam.cx) / cam.fx, (px[1] + 0.5 - cam.cy) / cam.fy, 1.0]
  dc /= norm(dc)
  # Rotate to world space
  R_c2w = cam_2_world(view)[:3, :3]
  # Compute ray direction
  rd = R_c2w @ dc
  return PixelInfoResult(rd=rd)

def evaluate(px_data, p_sd):
  ro_canonical, M, o, c = unpack(p_sd)
  rd = px_data[0:3]

  # Compute max response in canonical Gaussian space
  rd_canonical = M @ rd
  response = compute_max_response_canonical(ro_canonical,rd_canonical)

  alpha    = o * response  
  valid = (response > 0.0113)

  return EvaluateResult(alpha=alpha, valid=valid, color=c)

XPR_3DGUT = MethodSpec((*@project@*)=project, (*@tile\_cull@*)=tile_cull, (*@pixel\_info@*)=pixel_info, (*@evaluate@*)=evaluate)
\end{lstlisting}

%% file: sections/Implementation_XPRLinPrim.tex
\subsubsection{Implementation of \X-LinPrim}
\label{sec:listing_linprim}
\noindent\textbf{LinPrim}~\cite{von2025linprim} replaces 3D Gaussians with octahedra as the point-based scene representation. Each octahedron is defined by a center $\mu$, three half-extents $(s_x,s_y,s_z)$ along the local coordinate axes, and a rotation quaternion. Opposing vertices have equal distance from $\mu$, so the primitive can be described with three scale parameters rather than six (see Listing~\ref{lst:xprlinprim_params}).

\begin{lstlisting}[style=python, basicstyle=\scriptsize\ttfamily, caption={\texttt{XPR-LinPrim}: \texttt{PrimitiveParams}.}, label=lst:xprlinprim_params]
class PrimitiveParams:
  mu : Array  # (3,)  mean in world space
  s  : Array  # (3,)  half-extents (s_x, s_y, s_z)
  q  : Array  # (4,)  unit quaternion for rotation
  sh : Array  # (C,3) spherical harmonic coefficients
  o  : Array  # (1,)  opacity
\end{lstlisting}

Unlike Gaussians, octahedra have a homogeneous interior density:
\begin{equation*}
\sigma(o)=\frac{-\log(1-0.99\cdot o)}{2\cdot\min(s_x,s_y,s_z)} ,
\end{equation*}
and the per-pixel opacity follows the Beer-Lambert law,
\begin{equation*}
\alpha=1-\exp\!\left(-\sigma\,(\tau_{\mathrm{exit}}-\tau_{\mathrm{enter}})\right),
\end{equation*}
where $\tau_{\mathrm{enter}}$ and $\tau_{\mathrm{exit}}$ are the entry and exit depths along the per-pixel ray.

\X-LinPrim implements \texttt{project}, \texttt{tile\_cull}, and \texttt{evaluate} (see Listing~\ref{lst:xprlinprim}). Thus, like \texttt{\X-3DGS}, it sets \texttt{pixel\_info}{=}\texttt{None} and \texttt{evaluate} receives the pixel coordinate directly, applying the $+0.5$ pixel-center offset. The \texttt{project} function computes the camera-space center, pixel-space center, tile-space bounding box, Euclidean depth, homogeneous density, and the linear map from the local octahedron frame to the $(x_{\mathrm{px}},y_{\mathrm{px}},\tau)$ pixel-depth frame,
\begin{equation*}
A=J_{3\times3}R_{\mathrm{w2c},[:3,:3]}R\operatorname{diag}(s_x,s_y,s_z).
\end{equation*}
Here, $J_{3\times3}$ is the projective Jacobian augmented with a row that normalizes by Euclidean distance from the camera, $R_{\mathrm{w2c},[:3,:3]}$ is the rotation of the world-to-camera transform, and $R$ is the rotation of the primitive. The three rows $c_0,c_1,c_2$ of $A^{-1}$ define the projected octahedron as
\begin{equation*}
\{(p_{2D}, \tau):\|A^{-1}\begin{pmatrix}p_{2D}-\mu_{2D}\\ \tau \end{pmatrix}\|_1\le 1\}, 
\end{equation*}
where $\tau$ is relative to the center of the primitive. Equivalently, the projected octahedron is the intersection of four slab pairs with normals $c_0+c_1+c_2$, $c_0+c_1-c_2$, $c_0-c_1+c_2$, and $c_0-c_1-c_2$. The \texttt{evaluate} function intersects these slabs along the ray and converts the resulting chord length into opacity. The \texttt{tile\_cull} function applies the per-tile upper bound derived below. The diagonal-pair terms incorporate the primitive-level opacity criterion. Thus, no separate primitive-level opacity cull is required.

\input{listings/linprim}

\noindent\textit{Per-tile upper bound.} Fix a tile $t$ and let $\Omega_t=[x^{\min},x^{\max})\times[y^{\min},y^{\max})\subset \mathbb{R}^2$ denote the half-open rectangle in pixel space that contains the pixel centers of $t$. Index the four slab pairs by $i\in\{0,1,2,3\}$, with normals
\begin{equation*}
n_i=\varepsilon_{i,0}c_0+\varepsilon_{i,1}c_1+\varepsilon_{i,2}c_2,
\end{equation*}
where $\varepsilon_{i,0}=+1$ is fixed and $\varepsilon_{i,1}, \varepsilon_{i,2}\in\{+1,-1\}$. First assume $n_{i,z}\neq0$. Solving the inequality $|n_i\cdot(\Delta x,\Delta y,\tau)|\le1$ for $\tau$ gives
\begin{equation*}
s_i^{\mathrm{lo}}(x,y)=m_i(x,y)-h_i,\qquad
s_i^{\mathrm{hi}}(x,y)=m_i(x,y)+h_i,
\end{equation*}
where
\begin{equation*}
h_i=|n_{i,z}|^{-1}\ge0,\qquad
m_i(x,y)=-n_{i,z}^{-1}\bigl(n_{i,x}\Delta x+n_{i,y}\Delta y\bigr).
\end{equation*}
Here $m_i$ is affine in $(x,y)$, $\Delta x=x+\tfrac12-\mu_{2D,x}$, and $\Delta y=y+\tfrac12-\mu_{2D,y}$. The chord length through the intersection of the four slabs at pixel $(x,y)$ is
\begin{equation*}
\Delta \tau(x,y)=\max\!\left(0,\min_{i,j\in\{0,1,2,3\}}\left(s_i^{\mathrm{hi}}(x,y)-s_j^{\mathrm{lo}}(x,y)\right)\right),
\end{equation*}
using the identity $\min_i s_i^{\mathrm{hi}}-\max_j s_j^{\mathrm{lo}}=\min_{i,j}(s_i^{\mathrm{hi}}-s_j^{\mathrm{lo}})$.

We define $g_{ij}(x,y)=h_i+h_j+m_i(x,y)-m_j(x,y).$ Applying the weak minimax inequality and using the monotonicity of $\max(0,\cdot)$ gives
\begin{equation*}
\vcenter{\hbox{\footnotesize$\displaystyle\sup_{(x,y)\in \Omega_t}\Delta \tau(x,y)\le\max\!\left(0,\min_{i,j\in\{0,1,2,3\}}\left[h_i+h_j+\sup_{(x,y)\in \Omega_t}\left(m_i(x,y)-m_j(x,y)\right)\right]\right)$}}
\end{equation*}
Since $m_i-m_j$ is affine and continuous, its supremum over the half-open rectangle $\Omega_t$ equals its maximum over the closure $\overline{\Omega}_t$, and that maximum is attained at one of the four corners of $\overline{\Omega}_t$. Writing $C(\Omega_t)$ for the set consisting of the four corners of $\Omega_t$, the per-tile upper bound used by \texttt{tile\_cull} is
\begin{equation}
\vcenter{\hbox{\footnotesize $\displaystyle\Delta \tau_{\mathrm{upper}}(\Omega_t)=\max\!\left(0,\min_{i,j\in\{0,1,2,3\}}\left[h_i+h_j+\max_{c\in C(\Omega_t)}\left(m_i(c)-m_j(c)\right)\right]\right)$}}
\label{eq:linprim_tile_cull}
\end{equation}
The minimization is over $16$ ordered pairs because the chord differences $s_i^{\mathrm{hi}}-s_j^{\mathrm{lo}}$ and $s_j^{\mathrm{hi}}-s_i^{\mathrm{lo}}$ are distinct under the inner minimum. The diagonal pairs $i=j$ are included, and their entries reduce to $g_{ii}=2h_i$, which is independent of the pixel and the tile. Thus, only the twelve off-diagonal pair gaps require a maximum over the four corners in $C(\Omega_t)$. If $n_{i,z}=0$, the corresponding slab constraint is independent of $\tau$. For this upper bound, all pair gaps involving $i$ are treated as $+\infty$, equivalently removing that slab from the outer minimum. This can only increase $\Delta \tau_{\mathrm{upper}}(\Omega_t)$, so the bound remains conservative.

\noindent\textit{Cull predicate.} A primitive is removed from $t$'s candidate list whenever
\begin{equation*}
\sigma\,\Delta \tau_{\mathrm{upper}}(\Omega_t)<-\log(1-\alpha_{\mathrm{thresh}}).
\end{equation*}
The strict inequality matches the framework's non-strict per-pixel gate $\alpha\ge\alpha_{\mathrm{thresh}}$ at the boundary, so the primitive is kept when the bound reaches the threshold. Since $\sigma\ge0$ by construction and $\alpha=1-\exp(-\sigma\Delta \tau)$ is monotone non-decreasing in $\Delta \tau$, the chain
\begin{equation*}
\Delta \tau_{\mathrm{upper}}(\Omega_t)\ge\sup_{(u,v)\in \Omega_t}\Delta \tau(u,v)\ge\Delta \tau(x,y)
\end{equation*}
implies $\alpha(x,y)<\alpha_{\mathrm{thresh}}$ for every pixel of a culled primitive-tile pair. The framework's per-pixel alpha gate therefore masks such contributions to zero. In exact arithmetic, removing these entries is equivalent to keeping them with $\alpha=0$: the nonzero alpha sequence is unchanged, the transmittances $T_i=\prod_{k<i}(1-\alpha_k)$ are unchanged, and the composited color is unchanged. 

\noindent\textit{Redundancy of separate per-primitive opacity culling.}
The cases $i=j$ in Eq.~\ref{eq:linprim_tile_cull} reduce to $g_{ii}=2h_i$, which is independent of the pixel and tile. They therefore provide a global primitive-level bound on the chord length. Multiplying this bound by $\sigma$ gives the primitive-level bound used by the opacity test. Since the outer minimum in Eq.~\ref{eq:linprim_tile_cull} is taken over all ordered pairs, it cannot exceed the minimum obtained from only the cases $i=j$. By the monotonicity of $\max(0,\cdot)$, this gives
\begin{equation*}
\Delta \tau_{\mathrm{upper}}(\Omega_t)\le 2\min_i h_i .
\end{equation*}
Consequently, whenever the primitive-level bound satisfies
\begin{equation*}
2\sigma\min_i h_i<-\log(1-\alpha_{\mathrm{thresh}}),
\end{equation*}
every tile also satisfies the cull predicate. Thus, \X-LinPrim does not need a separate primitive-level opacity test. This matches the reference CUDA \texttt{linear-splatting} implementation~\cite{linprimbase_code}, which likewise uses frustum and tile-overlap culling without adding a separate primitive-level opacity test.

\noindent\textit{Comparison with StopThePop.}
Eq.~\ref{eq:linprim_tile_cull} follows the same algorithmic pattern as the per-tile Gaussian cull in StopThePop~\cite{radl2024stopthepop}: both methods compute a conservative tile-level upper bound on the maximum per-pixel opacity using the tile geometry and a small set of per-primitive constants, and both discard primitive-tile pairs whose bound is below the alpha threshold. The distinction comes from the opacity model. In LinPrim, each primitive has homogeneous density inside an $L_1$ polytope, so opacity is determined by the ray chord length through the primitive. The cull therefore bounds the longest possible chord over all pixels in $\Omega_t$. For a Gaussian, opacity decreases smoothly away from the mean, so StopThePop instead bounds the smallest quadratic distance from the tile to the Gaussian mean, equivalently the largest Gaussian response attainable within the tile. In both cases, the bound can become loose when the constraint attaining the maximum changes across $\Omega_t$, for example when different slab pairs or different closest points are active at different locations in the tile.

%% file: listings/linprim.tex
\begin{lstlisting}[style=python, basicstyle=\scriptsize\ttfamily, caption={Implementation of \texttt{\X-LinPrim}.}, label=lst:xprlinprim]
def project(p, cam, view, cfg):
  # Get world-to-camera projection matrix
  R_w2c = world_2_camera_projection_matrix(view)
  mu_c = world_2_cam(p.mu, R_w2c)
  mu_2D = compute_mu_2D(mu_c, cam)

  # Compute slab data, extent, radius, and depth
  c0, c1, c2, r, ex, ey, depth = compute_slab_data(p.s, p.q, mu_c, mu_2D, R_w2c, cam, cfg)

  # Compute bounding box
  aabb = get_bounding_box(mu_2D - 0.5, ex, ey, cfg)

  # Compute tile count
  tc = (aabb.x_max - aabb.x_min) * (aabb.y_max - aabb.y_min)

  # Compute density weight
  s_min = min(p.s[0], p.s[1], p.s[2])
  o = clip(p.o, 0.0, 0.999)
  sigma = -log1p(-0.99 * o) / (2.0 * s_min)

  # Compute visibility flag and adjust tile count
  visible = (r>0) & (tc>0) & (depth>cfg.z_near) & (s_min>0)
  tc *= float(visible)

  # Compute view-dependent color c
  c = compute_color(p.mu, p.sh, view)

  # Produce tile cull data and shader data for primitive p
  p_tcd = concat(mu_2D, c0, c1, c2, sigma, visible)
  p_sd = concat(mu_2D, c0, c1, c2, sigma, c)

  return ProjectResult(depth=depth, visible=visible, tile_cull_data=p_tcd, shader_data=p_sd, aabb=aabb, tile_count=tc)

def slab_m_at_corners(mu_2D, n, x0, y0, x1, y1):
  # Slab midpoint values at tile corners
  INF = 1e10
  EPS = 1e-10

  nz_ok = abs(n[2]) >= EPS
  nz = where(nz_ok, n[2], 1.0)
  inv_nz = where(nz_ok, 1.0 / nz, 0.0)

  corners = [(x0, y0), (x0, y1), (x1, y0), (x1, y1)]
  m_corner = [-inv_nz * (n[0] * (x - mu_2D[0]) + n[1] * (y - mu_2D[1])) for (x, y) in corners]
  half = where(nz_ok, abs(inv_nz), INF)

  return m_corner, half

def tile_cull(tile_min, tile_max, p_tcd, cfg):
  mu_2D, c0, c1, c2, sigma, visible = unpack(p_tcd)

  # Tile corners in pixel-center coordinates
  x0 = tile_min[0] + 0.5
  y0 = tile_min[1] + 0.5
  x1 = tile_max[0] + 0.5
  y1 = tile_max[1] + 0.5

  # Four slab-pair normals
  n[0] = c0 + c1 + c2
  n[1] = c0 + c1 - c2
  n[2] = c0 - c1 + c2
  n[3] = c0 - c1 - c2

  # Compute per-slab corner midpoints and half-widths
  for i in 0..3:
    m_corner[i], half[i] = slab_m_at_corners(mu_2D, n[i], x0, y0, x1, y1)

  # Compute 16 ordered pair gaps
  for i in 0..3:
    for j in 0..3:
      gap[i, j] = max_over_4_corners(m_corner[i] - m_corner[j]) + half[i] + half[j]

  # Bound the maximum chord length over the tile
  dt_upper = max(0.0, min_over_pairs(gap))

  # Keep tile when the opacity bound reaches the alpha threshold
  required = -log(1 - cfg.alpha_thresh)
  tile_visible = (sigma * dt_upper >= required) & visible
  return tile_visible

def evaluate(px_data, p_sd):
  mu_2D, c0, c1, c2, sigma, c = unpack(p_sd)

  # Pixel-center offset
  px = px_data[0:2] + 0.5

  # Ray-octahedron slab intersection
  tau_enter, tau_exit = slab_intersect(px, mu_2D, c0, c1, c2)

  # Beer-Lambert opacity
  ok = tau_enter < tau_exit
  dtau = where(ok, tau_exit - tau_enter, 0.0)
  alpha = where(ok, 1.0 - exp(-sigma * dtau), 0.0)

  return EvaluateResult(alpha=alpha, valid=1.0, color=c)

XPR_LinPrim = MethodSpec((*@project@*)=project, (*@tile\_cull@*)=tile_cull, (*@pixel\_info@*)=None, (*@evaluate@*)=evaluate)
\end{lstlisting}

%% file: tables/quality_comparison_appendix.tex
\begin{table}[!htb]
    \centering
    \setlength{\tabcolsep}{4pt}
    \resizebox{\columnwidth}{!}{
    \begin{tabular}{llccccccccc}
        \toprule
         & Dataset & \multicolumn{3}{c}{\texttt{360}} & \multicolumn{3}{c}{\texttt{T\&T}} & \multicolumn{3}{c}{\texttt{Synthetic}} \\
        \cmidrule(lr){3-5} \cmidrule(lr){6-8} \cmidrule(lr){9-11}
        Method & Metric & \texttt{Base} & \texttt{\X-4K} & \texttt{\X-8K}
                        & \texttt{Base} & \texttt{\X-4K} & \texttt{\X-8K}
                        & \texttt{Base} & \texttt{\X-4K} & \texttt{\X-8K} \\
        \midrule
        \multirow{2}{*}{\texttt{3DGS}}
            & SSIM  & 0.87   & 0.87   & 0.87   & 0.85   & 0.85   & 0.85   & 0.97   & 0.97   & 0.97   \\
            & LPIPS & 0.182  & 0.182  & 0.182  & 0.169  & 0.169  & 0.169  & 0.030  & 0.030  & 0.030  \\
        \midrule
        \multirow{2}{*}{\texttt{3DGUT}}
            & SSIM  & 0.87   & 0.87   & 0.87   & 0.86   & 0.85   & 0.85   & 0.97   & 0.97   & 0.97   \\
            & LPIPS & 0.184  & 0.184  & 0.184  & 0.168  & 0.169  & 0.168  & 0.030  & 0.030  & 0.030  \\
        \midrule
        \multirow{2}{*}{\texttt{LinPrim}}
            & SSIM  & 0.86   & 0.86   & 0.86   & 0.83   & 0.83   & 0.83   & 0.97   & 0.97   & 0.97   \\
            & LPIPS & 0.190  & 0.190  & 0.190  & 0.195  & 0.195  & 0.195  & 0.028  & 0.029  & 0.028  \\
        \bottomrule
    \end{tabular}
    }
    \caption{Comparison of average SSIM and LPIPS of \X with \texttt{Base} implementations for different point-based rendering methods and datasets on an \texttt{L40}.}
    \label{tab:quality_comparison_appendix}
\end{table}